%% file: lhapdf6-epjc.tex
\documentclass[twocolumn,epjc3]{svjour3}
\smartqed  % flush right qed marks, e.g. at end of proof
\RequirePackage{fix-cm}
\RequirePackage{graphicx}
\RequirePackage{xcolor}
\RequirePackage{xspace}
\RequirePackage{relsize}
\RequirePackage{amsmath}
\RequirePackage{url}
\RequirePackage{microtype}
\RequirePackage{booktabs}
\RequirePackage{textcomp}
\newcommand{\eV}{\ensuremath{\text{e}\mspace{-0.8mu}\text{V}\xspace}}

\newcommand{\GeV}{\ensuremath{\text{G\eV}}\xspace}
\newcommand{\TeV}{\ensuremath{\text{T\eV}}\xspace}

\newcommand{\alphaS}{\ensuremath{\alpha_\mathrm{S}}\xspace}
\newcommand{\LambdaQCD}{\ensuremath{\Lambda_\mathrm{QCD}}\xspace}

\newcommand{\kbd}[1]{{\smaller{\texttt{#1}}}}

\newcommand{\var}[1]{\textlangle\textit{#1}\textrangle\xspace}
\newcommand{\env}[1]{\kbd{\${#1}}\xspace}

\makeatletter
\g@addto@macro\bfseries{\boldmath}
\makeatother

\journalname{Eur. Phys. J. C}
\def\makeheadbox{\hfill GLAS-PPE/2014-05, MCnet-14-29, IPPP/14/111, DCPT/14/222
}
\begin{document}

\title{LHAPDF6: parton density access in the LHC precision era%\thanksref{t1}
}


%\author{First Author\thanksref{e1,addr1}
%        \and
%        Second Author\thanksref{e2,addr2,addr3} %etc.
%}

\author{Andy~Buckley\thanksref{e1,addr1} \and James~Ferrando\thanksref{addr1} \and Stephen~Lloyd\thanksref{addr2} \and Karl~Nordstr\"om\thanksref{addr1} \and Ben~Page\thanksref{addr3} \and Martin~R\"ufenacht\thanksref{addr4} \and Marek~Sch\"onherr\thanksref{addr5} \and Graeme~Watt\thanksref{addr6}}

\thankstext{e1}{e-mail: andy.buckley@cern.ch}


%\institute{First address \label{addr1}
%           Second address \label{addr2}
%           \emph{Present Address:} if needed\label{addr3}
%}

\institute{School of Physics \& Astronomy, University of Glasgow, UK \label{addr1}
  \and
  School of Physics \& Astronomy, University of Edinburgh, UK \label{addr2}
  \and
  Departamento de F\'isica Te\'orica y del Cosmos y CAFPE, Universidad de Granada, Spain \label{addr3}
  \and
  School of Informatics, University of Edinburgh, UK \label{addr4}
  \and
  Physik-Institut, Universit\"at Z\"urich, Switzerland \label{addr5}
  \and
  Institute for Particle Physics Phenomenology, Durham University, UK \label{addr6}
}

\date{Received: date / Accepted: date}

\maketitle

\begin{abstract}
  The Fortran LHAPDF library has been a long-term workhorse in particle physics,
  providing standardised access to parton density functions for experimental and
  phenomenological purposes alike, following on from the venerable PDFLIB
  package. During Run\,1 of the LHC, however, several fundamental limitations in
  LHAPDF's design have became deeply problematic, restricting the usability of
  the library for important physics-study procedures and providing dangerous
  avenues by which to silently obtain incorrect results.

  In this paper we present the LHAPDF\,6 library, a ground-up re-engineering of
  the PDFLIB/LHAPDF\linebreak[20] paradigm for PDF access which removes all
  limits on use of concurrent PDF sets, massively reduces static memory
  requirements, offers improved CPU performance, and fixes fundamental bugs in
  multi-set access to PDF metadata. The new design, restricted for now to
  interpolated PDFs, uses centralised numerical routines and a powerful
  cascading metadata system to decouple software releases from provision of new
  PDF data and allow completely general parton content. More than 200 PDF sets
  have been migrated from LHAPDF\,5 to the new universal data format, via a
  stringent quality control procedure. LHAPDF\,6 is supported by many Monte
  Carlo generators and other physics programs, in some cases via a full set of
  compatibility routines, and is recommended for the demanding PDF access needs
  of LHC Run\,2 and beyond.

  % \keywords{First keyword \and Second keyword \and More}
\end{abstract}

\tableofcontents

\section{Introduction}
\label{sec:intro}
\input{intro}

\section{History and evolution of LHAPDF}
\label{sec:history}
\input{history}

\section{Design of LHAPDF\,6}
\label{sec:design}
\input{design}

\section{Usage examples}
\label{sec:usage}
\input{usage}

\section{Data formats}
\label{sec:data}
\input{data}

\section{PDF uncertainties}
\label{sec:uncertainties}
\input{uncertainties}

\section{PDF reweighting}
\label{sec:reweighting}
\input{reweighting}

\section{LHAPDF\,5 / PDFLIB compatibility}
\label{sec:compatibility}
\input{compatibility}

\section{Benchmarking and performance}
\label{sec:performance}
\input{performance}

\section{PDF migration and validation}
\label{sec:migration}
\input{migration}

\section{Summary and prospects}
\label{sec:summary}
\input{summary}

\begin{acknowledgements}
  Thanks to Jeppe~Andersen, Juan~Rojo, Luigi~del~Debbio, Richard~Ball, and
  Nathan~Hartland for helpful suggestions and inputs on PDF collaboration
  requirements, which were invaluable in evolving this design. Many thanks also
  to David Hall, who provided the \kbd{lhapdf} data management script, to
  David~Mallows for early help with the interpolator code and Python interface,
  and to Gavin~Salam for several suggestions and a fast numeric ASCII parser code.

  AB wishes to acknowledge support from a Royal Society University Research
  Fellowship, a CERN Scientific Associateship, and IPPP Associateships during
  the period of LHAPDF\,6 development. IPPP grants also supported the work of
  SL, MR, and David Mallows on this project. KN thanks the University of Glasgow
  College of Science \& Engineering for a PhD studentship scholarship.
\end{acknowledgements}


%\bibliography{lhapdf6-epjc}   % name your BibTeX data base

\bibliographystyle{JHEP}       % APS-like style for physics
\bibliography{lhapdf6-jhep}   % name your BibTeX data base

% % Non-BibTeX users please use
% \begin{thebibliography}{}
% %
% % and use \bibitem to create references. Consult the Instructions
% % for authors for reference list style.
% % Format for Journal Reference
% Author, Article title, Journal, Volume, page numbers (year)
% Author, Book title, page numbers. Publisher, place (year)
% \end{thebibliography}

\end{document}

%% file: intro.tex
%auto-ignore

Parton density functions (PDFs) are a crucial input into cross-section
calculations at hadron colliders; they encode the process-independent momentum
structure of partons within hadrons, with which partonic cross-sections must be
convolved to obtain physical results that can be compared to experimental data.
At leading order in perturbation theory, PDFs encode the probability that a beam
hadron's momentum is carried by a parton of given flavour and momentum
fraction. At higher orders this interpretation breaks down and positivity is no
longer required -- but PDF normalization at all orders is constrained by the
requirement that a sum over all parton flavours $i$ and momentum fractions $x$
equates to the whole momentum of the incoming beam hadron $B$:
\begin{equation}
  \label{eq:xsumrule}
  \sum_i \int_0^1 \! \mathrm{d}x \; x \, f_{i/B}(x; Q^2) = 1,
\end{equation}
where $f_{i/B}(x; Q^2)$ is the parton density function for parton $i$ in $B$, at a
factorization scale $Q$. Conservation of baryon number leads to a
flavour sum rule,
\begin{equation}
  \label{eq:flsumrule}
  \int_0^1 \! \mathrm{d}x \; \left( f_{i/B}(x; Q^2) - \bar{f}_{i/B}(x; Q^2) \right) = n_i,
\end{equation}
where $i$ runs over quark flavours and $\bar{f}_{i/B}$ is the antiquark PDF in
baryon $B$. For protons, $n_u = 2$, $n_d = 1$, and $n_{\{s,c,b,t\}} = 0$.

Parton density calculations sit astride the borderline of perturbative and
non-perturbative QCD, constructed by fitting of a factorised low-scale,
non-\-perturb\-ative component to experimental data and then evolved to higher
scales using perturbative QCD running, most commonly DGLAP evolution. In
general, PDFs may include a transverse momentum dependence %\,\cite{tmdpdfs}
but here % and in the LHAPDF library
we restrict ourselves to collinear PDFs where the extracted parton momenta are
perfectly aligned with that of the parent hadron; such PDFs are then defined as
a two-variable function $f_{i/B}(x; Q)$ for collinear momentum fraction $x$ and
factorization scale $Q$. Eqs.~\eqref{eq:xsumrule} and \eqref{eq:flsumrule} apply
independently at each value of $Q$, hence the semicolon separator between $f$'s
parameters.

The LHAPDF library is the ubiquitous means by which parton density functions are
accessed for LHC experimental and phenomenological studies. It is both a
framework for uniform access to the results of many different PDF fitting groups
and a collection of such PDF sets. The first version of LHAPDF was developed to
solve scaling problems with the previously standard PDFLIB
library\,\cite{PlothowBesch:1992qj}, and to retain backward compatibility with it; in this
paper we describe a similar evolution within the LHAPDF package, from a
Fortran-based static memory paradigm to a C++ one in which dynamic PDF object
creation, concurrent usage, and removal of artificial limitations are
fundamental. This new version addresses the most serious limitations of the
Fortran version, permitting a new level complexity of PDF systematics estimation
for precision physics studies at the LHC\,\cite{Bruning:2004ej} Run\,2 and beyond.

\subsection{Definitions and conventions}
\label{defns}

Since the beam hadron will in most current applications be a proton, we will
simplify the notation from here by dropping the $/B$ specification of the parent
hadron, i.e. $f_i(x; Q^2)$ rather than $f_{i/B}(x; Q^2)$. Other parent hadrons
are possible, of course, notably neutrons which can either be fitted explicitly
or obtained from proton PDFs assuming strong isospin symmetry.

The PDFs appear in hadron collider cross-section
calculations in the form~\cite{Campbell:2006wx,Forte:2013wc}:
\begin{equation}
  \label{eq:xsec}
  \sigma =
  %\sum_{i,j}
  % \int_0^1 \!\mathrm{d}x_1 \int_0^1 \!\mathrm{d}x_2 \;
  \int \! \mathrm{d}x_1 \mathrm{d}x_2 \;
  f_i(x_1; Q^2) \; f_j(x_2; Q^2) \; \hat{\sigma}_{ij}(x_1, x_2, Q^2) ,
\end{equation}
where $\hat{\sigma}_{ij}$ is the partonic cross-section for a process with
incoming partons $i$ and $j$. Usually several partonic initial states contribute
and should be summed over in Eq.~\eqref{eq:xsec}.

Given the fundamental role played by the $x f(x; Q^2)$ structure in the fitting and
use of PDFs, it is this form which is encoded in the LHAPDF library. We will
tend to refer to this encoded value as the ``PDF value'' or similar, even though
it is in fact a combination of the parton density function and the momentum
fraction $x$.

Another ambiguity in common usage is the meaning of the words ``PDF set'', which
are sometimes used interchangeably with ``PDF'' and sometimes not. If one
considers a PDF to be a function defined for a given parton flavour, then both a
collection of such functions for all flavours, and a larger collection of
systematic variations on such collections can reasonably be called a ``PDF
set''. In this paper, particularly when referring to LHAPDF code objects, we
will take the approach that a ``PDF'' or ``PDF set member'' is a complete set of
1-flavour parton density functions; we refer to a larger collection of
systematic variations on such an object, e.g.~eigenvectors or Monte Carlo (MC) replicas, as a
``PDF set''.

Finally, when referring to code objects or configuration directives we will do
so in \kbd{typewriter font}. %, for clarity of meaning.

%% file: history.tex
%auto-ignore

LHAPDF versions 5 and earlier\,\cite{Whalley:2005nh,Bourilkov:2006cj} arose out
of the 2001 Les Houches ``Physics at \TeV Colliders''
workshop\,\cite{Giele:2002hx}, as the need for a scalable system to replace
PDFLIB became pressing. The main problem with\linebreak[20] PDFLIB was that the
data for interpolating each PDF was stored in the library, and as PDF fitting
became industrialised (particularly with the rise of the CTEQ and MRST error
sets), this model was no longer viable. %due to the explosion in the size of
% the compiled library.

LHAPDF was originally intended to address this problem by instead storing only
the parameters of each parton density fit at a fixed low scale and then using
standard DGLAP evolution in $Q$ via QCDNUM\,\cite{Botje:2010ay} to dynamically
build an interpolation grid to higher scales, and thereafter work as
before. However, by the mid-2000s and version 4 of LHAPDF, this model had also
broken down. Each PDF parameterisation required custom code to be included in
the LHAPDF library, and the bundled QCDNUM within LHAPDF had itself become
significantly outdated: upgrading it was not an option due to the need for
consistent behaviour between LHAPDF versions. PDF fitting groups, concerned that
the built-in QCDNUM evolution would not precisely match that used by their own
fitting code, universally chose to supply full interpolation grid files
rather than evolution starting conditions, and as a result LHAPDF acquired a large
collection of routines to read and use these data files in a myriad of formats.
% At the time of writing, of the many actively used PDFs in use for LHC simulation
% and phenomenology only the CTEQ6L1\,\cite{Pumplin:2002vw} PDF uses QCDNUM
% evolution; all others are interpolation-based.

At the same time as these trends back to interpolation-based PDF provision, user
demand resulted in new features for simultaneous use of several PDF sets -- the
so-called ``multiset'' mode introduced in LHAPDF\,5.0. The implementation of
this was relatively trivial: the amount of allocated interpolation space was
multiplied by a factor of \kbd{NMXSET} (with a default value of 3), but while it
permitted rapid switching between a few concurrent sets the multiset mode did
not integrate seamlessly with the original interface, potentially leading to
incorrect results, and was memory-inefficient and limited in
scalability. % Indeed, to calm the rise
% in memory requirements linked to multiset mode and increasing numbers of PDF
% interpolation routines, a ``low memory'' mode was introduced in which only space
% for a single member of each PDF set would be allocated at one time.

% Hence in a default LHAPDF\,5
% build, it is possible to switch rapidly between up to three PDF sets (although
% only one of these ``slots'' can be active at any time).

\subsection{Performance problems}

The major problems with LHAPDF v5 relate to the technical implementation of the
various interpolation routines and the multiset mode.

Both these issues are rooted in Fortran's static memory allocation. As usual,
the interpolation routines for various PDFs operate on large arrays of floating
point data. These were typically declared as Fortran common blocks, but in
practice were not used commonly: each PDF group's ``wrapper'' code operates on
its own array. As the collection of supported PDF sets became larger, the memory
requirements of LHAPDF continually grew, and with version 5.9.1 (the final
version in the v5 series) more than 2\,GB was declared as necessary to use it at
all. In practice operating systems did not allocate the majority of this
uninitialised memory, but it proved a major issue for use of LHAPDF on the LHC Computing
Grid system where static memory restrictions had to be passed in order for a job
to run.

 % In
% practice this is not a ``real'' issue -- operating systems are not so foolish as
% to allocate memory to store uninitialized data for PDFs which are not being used
% -- but on memory managed systems such as the LHC Grid where 2\,GB is the maximum
% level permitted per core/process, it is enough to block any LHAPDF\,5 jobs from
% running.

A workaround solution was provided for this problem: a so-called ``low memory''
build-time configuration which reduced the static memory footprint within
acceptable limits, but at the heavy cost of only providing interpolation array
space for one member in each PDF set.  This mode is usually sufficient for event
generation, in which only a single PDF is used, and in this form it was used for
the LHC experimental collaborations' MC sample production through LHC
Run\,1. But it is incompatible with ``advanced'' PDF uncertainty studies in
which each event must be re-evaluated or reweighted to every member in the PDF
error set: constant re-initialisation of the single PDF slots from the data file
slows operations to a crawl. For this reason, and because the low-memory mode is
a build-time rather than run-time option, PDF reweighting studies for the LHC
needed to use special, often private, user builds of LHAPDF with the attendant
danger of inconsistency.

The era of the low-memory mode's suitability for event generation has also come
to an end between LHC Runs 1 and 2, with the rise of next-to-leading order (NLO)
matrix element calculations ``matched'' to parton shower
algorithms\,\cite{Frixione:2002ik,Frixione:2007vw}. The ``NLO revolution'' has been a great
success of LHC-era phenomenology and the bulk of Standard Model %(and Higgs)
processes are now simulated at fully-exclusive NLO -- but the flip-side is that
PDF reweightings now require detailed information about initial parton
configurations in each NLO subtraction
counter-term\,\cite{Frederix:2011ss}. Accordingly PDF uncertainties are
increasingly calculated as event weightings \emph{during} the generation rather
than retrospectively as done in the past for leading-order (LO)
processes.\footnote{NLO event generators may report summary PDF information, for
  example in HepMC's \kbd{PdfInfo} object, but this is an approximation and may
  give very misleading effects if used for retrospective reweighting.}

% Notably, low-memory mode is incompatible with the recent trend to compute (in
% particular NLO) PDF reweightings within an MC generator run rather than post
% hoc, since every event involves upwards of 40 slow PDF re-initializations which
% dominate the generation time: large scale experiment MC production with this
% feature enabled in aMC@NLO\,\cite{Hirschi:2011pa},
% Sherpa\,\cite{Gleisberg:2008ta}, or POWHEG\,\cite{Alioli:2010xd} requires an
% improved LHAPDF.

Further options exist for selective disabling of\linebreak[20] LHAPDF support
for particular PDF families, as an alternative way to reduce the memory
footprint. However, since this highly restricts the parton density fits which
can be used, it has not found much favour.

Of course, with a design so dependent on global state and shared memory, Fortran
LHAPDF is entirely unsafe for use in multi-threaded applications: this greatly
restricts its scalability in the current multi-core computing era.

\subsection{Correctness problems}

The last set of problems with LHAPDF\,5 relate, concerningly, to the
\emph{correctness} of the output. For example different generations of PDF fit
families share the same interpolation code, although they may have different
ranges of validity in $x$--$Q$ phase space, and wrong ranges are sometimes
reported.

The reporting of $\Lambda_\text{QCD}$ and other metadata has also been
problematic, to the extent that PYTHIA\,6's many tunes depend on LHAPDF
returning a nonsense value which is then reset to the default of 0.192\;\GeV.
Since the multiset mode is often only implemented as a multiplying factor on the size
and indexing offsets, reported values of metadata such as \alphaS and $x$ \& $Q$
boundaries do not always correspond to the currently active PDF slot, but
rather to properties of the last set to have been initialised.

\subsection{Maintainability problems}

Aside from the technical issues discussed above, the design of LHAPDF\,5 (and
earlier versions) tightly couples PDF availability to the release cycle of the LHAPDF code
library -- as in PDFLIB. As PDF fitting has become more diverse, with many
different groups releasing PDF fits in response to new LHC and other data, the
mismatch of the slow software releases (typically two releases per year) and the
faster, less predictable release rate of new PDF sets has become evident. It is
neither desirable for new PDF data to have to wait for months before becoming
publicly available via an LHAPDF release, nor for experiments and other users to
be deluged with new software versions to be installed and tested.

In addition, since adding new PDFs involved interfacing external Fortran code
via ``wrapper'' routines, it both required significant coding and testing work
from the LHAPDF maintainers, and blocked PDF fitting groups from using languages
other than Fortran for their fitting/interpolation codes. The (partial) sharing
of wrapper routines between some sets which did not provide their own
interpolation code made any changes to existing wrapper code dangerous and
fragile. An attempt was made to make it easier for users to make custom PDFs by
using one of three generic set names to trigger a polynomial spline
interpolation, but this was also very restricted in functionality and saw
minimal use.

A final logistical issue was the lack of version tracking in PDF data files,
which would periodically be found to be buggy, and no way to indicate which
versions of the LHAPDF library were required to use a particular PDF. This led
to some problems where for space-saving reasons PDF data would be shared between
different versions of the library, producing unintended numerical changes and
potentially introducing buggy outputs from previously functional installations.

\subsection{Summary of LHAPDF\,5 issues}

Many of the problems of LHAPDF\,5 stem from the combination of the static nature
of Fortran memory handling and from the way that evolving user demands on LHAPDF
forced retro-fitting of features such as grid interpolation and multiset mode on
to a system not originally designed to incorporate them. These have combined
with more logistical features such as the lack of any versioned connection
between the PDF data files and the library, the menagerie of interpolation grid
formats, and the need to modify the library to use a new PDF to make LHAPDF\,5
difficult both to use and to maintain. These issues became critical during
Run\,1 of the LHC, leading to the development of LHAPDF\,6 to deal with the
increased demands on parton density usage in Run\,2 and beyond. Version 5.9.1 of
LHAPDF was the last in the Fortran series; all new development and maintenance
(including provision of new PDF sets) is restricted to LHAPDF\,6 only.

%% file: design.tex
%auto-ignore

LHAPDF\;6 is a ground-up redesign and re-implement\-ation of the LHAPDF system,
specifically to address all the above problems of the Fortran LHAPDF versions.
As so many of these problems fundamentally stem from Fortran(77) static memory
limitations, and the bulk of new experimental and event generator code is
written in C++, we have also chosen to write the new LHAPDF\,6 in object
oriented C++. Since the Python scripting language has also become widely used in
high-energy physics, we also provide a Python interface to the C++ LHAPDF
library, which can be particularly useful for interactive PDF testing and
exploration.

\subsection{PDF value access}

The central code/design object in LHAPDF\,6 is the \kbd{PDF}, an interface class
representing parton density functions for several parton flavours, typically but
not necessarily the gluon plus the lightest 5 quark (and anti-quark) flavours. An
extra object, \kbd{PDFSet} is provided purely for (significant) convenience in
accessing PDF set metadata and all the members in the set, e.g. for making
systematic variations within a set. The set level of data grouping is
unavoidable, even in the case of single-member sets, and a list of all available
PDF sets on the user's system can be obtained via the
\kbd{LHAPDF::availablePDFSets()} function. There is no LHAPDF\,6 user-interface
type to represent a single-flavour parton density.

Unlike in LHAPDF\,5, where a few PDFs included a parton density for a
non-standard flavour such as a photon or gluino via a special-case
``hack''\,\cite{Martin:2004dh,Berger:2004mj}, LHAPDF\,6 allows completely
general flavours, identified using the standard PDG Monte Carlo~ID
code\,\cite{Beringer:1900zz} scheme. An alias of 0 for 21 = gluon is also
supported, for backward compatibility and the convenience of being able to
access all QCD partons with a for-loop from -6 to 6.
% A hypothetical \kbd{PDF}
% might hence declare, for example, that it only contains explicit up, down and
% gluino flavours, and all other parton flavours would return $xf(x,Q^2) = 0$
% everywhere; while this would not describe data, there is no technical obstacle
% to its construction.

$xf(x; Q^2)$ values are accessed via the \kbd{PDF} interface methods
\kbd{PDF::xfxQ(...)} and \kbd{PDF::xfxQ2(...)} -- the only distinction between
these name variants is whether the scale argument is provided as an energy or
energy-squared quantity. The most efficient way is the $Q^2$ argument, since
this is the internal representation -- it is more efficient to square a $Q$
argument than to square-root a $Q^2$ one. Overloadings of these functions'
argument lists allow PDF values to be retrieved from the library either for a
single flavour at a time, for all flavours simultaneously as a \kbd{int} $\to$
\kbd{double} \kbd{std::map}, or for the standard QCD partons as a (pre-existing)
\kbd{std::vector} of \kbd{double}s. Parton flavours not explicitly declared in a
\kbd{PDF} object will return $xf(x; Q^2) = 0$.

\subsection{PDF metadata}

A key feature in the LHAPDF\,6 design is a powerful ``cascading metadata''
system, whereby any information (integer, floating point, string, or homogeneous
lists of them) can be attached to a \kbd{PDF}, a \kbd{PDFSet}, or the global
configuration of the LHAPDF system via a string-valued lookup key.
Access to metadata is via the general \kbd{Info} class, which is used directly
for the global LHAPDF system configuration and specialised into the \kbd{PDFSet}
and \kbd{PDFInfo} classes for set-level and PDF-level metadata respectively.

Much of the physics content of LHAPDF is in fact encoded via the metadata system.
% -- the numerical core of the library is not very specific to parton densities,
% and has actually already been used for other purposes such as fragmentation
% functions\,\cite{GenEVA}.
For example, the value of $\alphaS(M_Z)$ is defined via metadata: if it is not
defined on a \kbd{PDF}, the system will automatically fall back to looking in
the containing \kbd{PDFSet}, and finally the LHAPDF configuration for a value
before throwing an error (or accepting a user-supplied default). The metadata
information is set in the PDF/PDF set/global configuration data files, as
described later, and any metadata key may be specified at any level (with more
specific levels overriding more generic ones). The main motivation for the
cascade is reduced duplication and easier configuration: a global change in
behaviour need not be set in every \kbd{PDF}, and set-level information need not
be duplicated in the data files for every one of its members. All metadata
values set from file may also be explicitly overridden in the user code.

\subsection{Object and memory management}

A very important change in LHAPDF\,6 with respect to v5 is how the user manages
the memory associated with PDFs -- namely that they are now fully responsible
for it. A user may create as many or as few \kbd{PDF}s at runtime as they wish
-- there is neither a necessity to create a whole set at a time, nor any need to
re-initialise objects, nor a limitation to \kbd{NMXSET} concurrent PDF
sets. The flip-side to this flexibility is that the user is also responsible for
cleaning up this memory use afterwards, either with manual calls to \kbd{delete}
or by use of e.g. smart pointers.

Many objects, including \kbd{PDFs}, are created in factory functions such as
\kbd{LHAPDF::mkPDF(...)}, \kbd{LHAPDF::getPDFSet\-(...)}, and
\kbd{LHAPDF::PDFSet::mkPDFs()}. Internally these functions typically call the
\kbd{new} operator so that the memory is allocated on the heap and outlives the
scope of the calling function. We use a naming convention to indicate when the
user needs to delete the created objects: if the function name starts with
``\kbd{mk}'', then the return type will be pointer(s) and the user is
responsible for deletion. Note that \kbd{LHAPDF::getPDFSet(...)} is \emph{not}
such a function: \kbd{PDFSet} is a lightweight object shared between the set
members and hence its memory is automatically managed and is only exposed to the
user via a reference handle, not a pointer.

Creation of PDFs is usually done via the factory functions
\kbd{LHAPDF::mkPDF(...)} and \kbd{LHAPDF::mkPDFs(...)},\linebreak[20] which take
several forms of argument list. \kbd{mkPDF}, which returns a heap-allocated
\kbd{PDF*}, either takes two identifier arguments -- the string name of the PDF
set, plus the integer PDF member offset within the set -- or a single string
which encodes both properties with a slash separator,
e.g. \kbd{mkPDF("CT10nlo/0")} to refer to the central member of the CT10nlo
set. For convenience, if the \kbd{/0} is omitted when specifying a single PDF,
the first (nominal) member is taken as implied. This string-based lookup is
extremely convenient\footnote{Extension of this scheme is anticipated for PDFs
  with nuclear correction factors in a future release.} and we encourage uptake
of this scheme as standard syntax for referencing individual PDF members. A
final form takes a single integer argument, which gives the global LHAPDF ID
code for the desired PDF set member.  The \kbd{mkPDFs(...)} functions behave
similarly, but only the set name is specified (or implied when calling
\kbd{LHAPDF::PDFSet::mkPDFs()}). If no further argument is given, the PDFs are
returned as a \kbd{vector<PDF*>}, but an extra argument of templated type
\kbd{vector<T>} may also be given and will be filled in-place for better
computational efficiency and to allow automatic use of smart pointers.

\subsection{PDF value calculation}

The PDF $xf(x; Q^2)$ values may come from any implementation, derived from the
abstract \kbd{PDF} class, although (reflecting the reality of real-world PDF
usage) the only current provider is the \kbd{GridPDF} class which provides PDF
values interpolated from data files.%  of values at points on 2D grids of $(x,
% Q^2)$ coordinates.

% Following the strong trend towards interpolated PDFs, whereby PDF groups can
% obtain arbitrarily faithful representations of their fits by increasing the
% sampling density, and away from internally QCD-evolved PDFs, the only internally
% provided PDF type in LHAPDF\,6 is an interpolation on a rectangular grid in
% $x$--$Q^2$ space.

These data files consist of PDF values for each flavour evaluated on a rectangular
grid of ``knots'' in $(x, Q^2)$, with values for all flavours given at
each point. The spacing of the knot positions in $x$ and $Q^2$ is not
prescribed, but the physical nature of PDFs means that most natural and
efficient representation is to use uniform or near-uniform distributions in
$\log x$ and $\log Q^2$.

In fact, each PDF may contain arbitrarily many distinct grids in $Q^2$, in order
to allow for parton density discontinuities (or discontinuous gradients) across
quark mass thresholds. This gives the possibility of correct handling of
evolution discontinuities in NNLO PDFs, and is used by the MSTW2008 and
NNPDF\,3.0 fits. There is no requirement that the subgrid boundaries lie on quark
masses -- they may be treated as more general thresholds if wished. The $Q^2$
boundaries of these subgrids, and the $x$, $Q^2$ knots within them must be the
same for all flavours in the PDF. The mechanisms for efficient lookup from an
arbitrary $(x, Q^2)$ to the containing subgrid, and of the surrounding knots
within that subgrid (and of specific flavours at each point) are implemented in
the \kbd{GridPDF} class and associated helper structures.

Since several applications of PDFs, notably their use in Monte Carlo parton
shower programs, require a probabilistic interpretation of the PDF values, a
``force positive'' option has been implemented to ensure (if requested) that
negative $xf(x;Q^2)$ values are not returned, either from actual negative values
at interpolation knots or by a vagary of the interpolation algorithm. This is
necessary for leading-order or leading-log applications such as parton showers,
but not in the matrix element computation of NLO shower-matched generators. The
force-positive behaviour is set via the \kbd{ForcePositive} metadata key,
which takes values of 0, 1, or 2 to, respectively, indicate no forcing, forcing
negative values to 0, or forcing negative-or-zero values to a very small
positive constant.

The interpolation of gridded PDF values to arbitrary points within the grid $x$
and $Q^2$ ranges is performed by a flexible system of interpolator objects.

\subsubsection{Interpolator system}
\label{sec:interpolation}

There are many possible schemes for PDF interpolation. To strike a balance
between efficiency and complexity, we have implemented an interpolation based on
cubic Hermite splines in $\log Q^2$--$\log x$ space as the default interpolation
scheme in LHAPDF\,6, implemented in the \kbd{LogBicubicInterpolator} class,
which inherits from an abstract \kbd{Interpolator} type.

Internally, the log-cubic PDF querying is natively done via $Q^2$ rather than
$Q$, since event generator shower evolution naturally occurs in a squared energy
(or $p_\perp$) variable and it is advisable to minimise expensive calls of
\kbd{sqrt}. For this log-based interpolation measure, the logarithms of (squared)
knot positions are pre-computed in the interpolator construction to avoid
excessive \kbd{log} calls in calls to the interpolation function. In the regions
close to the edges of each subgrid, where fewer than the minimum number of knots
required for cubic spline interpolation are available, the interpolator switches
automatically to linear interpolation.

This interpolation scheme is not hard-coded but is simply the standard value,
``logcubic'', of the \kbd{Interpolator} metadata key. This key is read at
runtime when a PDF's value is first queried, and is used as the argument to a
factory function whose job is to return an object implementing the
\kbd{Interpolator} interface. If an alternative value is specified in the PDF
set's \kbd{.info} file, in a specific member's \kbd{.dat} file, or is overridden
by a call to \kbd{PDF::setInterpolator(...)} before the PDF is first queried,
then the corresponding interpolator will be used instead. At present, however,
the alternative interpolators such as ``linear'' are intended more for debugging
(and for edge-case fallbacks) than for serious physics purposes.

As the interpolator algorithm is runtime-configurable, there is the possibility
of evolving better interpolators in a controlled way without changing previous
PDF behaviours. So far there has been little incentive to do so, as specific
problem regions like high-$x$ where uniform spacing of anchor points in $\log x$
becomes sub-optimal are most easily dealt with by locally increased knot density
rather than a global increase in the complexity (and computational cost) of the
interpolation measure.

Interpolation as described here only applies within the limiting ranges of the
$(x, Q^2)$ grid (given by \kbd{XMin}--\kbd{XMax} and \kbd{QMin}--\kbd{QMax}
metadata keys and accessed most conveniently via the \kbd{PDF::xMin()}
etc. methods). Outside this range, a similar extrapolator system is used.

\subsubsection{Extrapolation system}
\label{sec:extrapolation}

The majority of PDF interpolation codes included in LHAPDF\,5 did not return a
sensible extrapolation outside the limits of the grid, with many codes even
returning nonsensical PDF values.  Hence the default LHAPDF\,5 behaviour was to
``freeze'' the PDFs at the boundaries, although this option could be overridden
for the few PDF sets that did return sensible behaviour beyond the grid limits.

In particular, the MSTW interpolation code included in LHAPDF\,5 made an effort
to provide a sensible extrapolation to small-$x$, low-$Q$ and high-$Q$ values.
A continuation to small $x$ values was performed by linearly extrapolating from
the two smallest $\log x$ knots either the value of $\log xf$, if $xf$ was
sufficiently positive, or just $xf$ itself otherwise.  A similar continuation to
high $Q$ values was performed based on linear extrapolation from the two highest
$\log Q^2$ knots.  Extrapolation to low $Q$ values is more ambiguous, but the
choice made was to interpolate the anomalous dimension, $\gamma(Q^2)
= \partial\log xf/\partial\log Q^2$, between the value at $Q_\mathrm{min}$ and a
value of $1$ for $Q\ll Q_\mathrm{min}$, so that the PDFs for $Q \sim
Q_\mathrm{min}$ behave as:
\begin{equation}
xf(x;Q^2)=xf(x;Q_\mathrm{min}^2)\left(Q^2/Q_\mathrm{min}^2\right)^{\gamma\left(Q_\mathrm{min}^2\right)},
\end{equation}
while for $Q\ll Q_\mathrm{min}$ the PDFs vanish as $Q^2\to 0$ like:
\begin{equation}
xf(x;Q^2)=xf(x;Q_\mathrm{min}^2)\left(Q^2/Q_\mathrm{min}^2\right).
\end{equation}

In LHAPDF\,6, $(x, Q^2)$ points outside the grid range trigger the same sort of
function-object lookup as for in-range interpolation, but the returned object
now implements the \kbd{Extrapolator} interface.

The default extrapolation, as of LHAPDF version 6.1.5, is an implementation of
the MSTW scheme for use with all PDF sets, named the ``continuation''
extrapolator. Alternatives are also available: a ``nearest'' extrapolator as in
LHAPDF\,5, which operates by identifying the nearest in-range point in the grid
and then using the correct interpolator to return the value at that point via a
pointer back to the \kbd{GridPDF} object; and an ``error'' extrapolator which
simply throws an error if out-of-range PDF values are requested. Uncontrolled
evolution outside the range is not an option for LHAPDF\,6's interpolation
grids.

% TODO: plot showing e.g. gluon distribution with continuation extrapolation to low-x, low-Q and high-Q compared to "frozen" behaviour.

\subsection{$\alphaS$ system}
\label{sec:alphas}

Consistent \alphaS evolution is key to correct PDF evolution and usage: programs
which use PDFs in cross-section calculations should also ensure, at least within
fixed-order perturbative QCD computations, that they use \alphaS values
consistent with those used in the PDF fit. LHAPDF\,6 contains implementations of
\alphaS running via three methods: an analytic approximation, a numerical
solution of the ODE, and a 1D cubic spline interpolation in $\log Q$. All three
methods implement the \kbd{LHAPDF::AlphaS} interface.

The first two of these methods are defined within the $\overline{\text{MS}}$
renormalization-scheme, and for consistency this scheme should also be used for
interpolation values supplied to the spline interpolation. The analytic and ODE
implementations are based on the outlines given in\linebreak[20]
Ref.~\cite{Beringer:1900zz} using the result from
Ref.~\cite{vanRitbergen:1997va} for $b_3$, the results from
Ref.~\cite{Chetyrkin:2005ia} for the QCD decoupling coefficients $c_n$, and the
result from Ref.~\cite{PhysRevLett.79.2184} for the analytic four-loop
approximation.  Flavour thresholds/masses, orders of QCD running, and fixed
points/\LambdaQCD are all correctly handled in the analytic and ODE solvers, and
subgrids are available in the interpolation.

\paragraph{The ODE solver} approximates the \alphaS running by numerically solving the
renormalization group equation up to four-loop order using the input parameters
$M_Z$, $\alphaS(M_Z)$:
\begin{align}
 \mu^2 \frac{d \alphaS}{d \mu^2}
 &= \beta(\alphaS)\\
 &= - \left( b_0 \alphaS^2 + b_1 \alphaS^3 + b_2 \alphaS^4 + b_3 \alphaS^5 + O(\alphaS^6) \right).
\end{align}
The decoupling at flavour thresholds where we go from $n_f$ to $n_f + 1$ active
flavours or vice versa is currently calculated using under the assumption the
flavour threshold is at the heavy quark mass, a restriction which will shortly
be relaxed to allow use of generalised thresholds:
\begin{equation}
 \alphaS^{(n_f + 1)}(\mu) = \alphaS^{(n_f)}(\mu) \left( 1 + \sum_{n=2}^{\infty} c_n [\alphaS^{(n_f)}(\mu)]^n \right).
\end{equation}
If a more involved calculation is required, we suggest linking LHAPDF6 to a
dedicated \alphaS library such as that described in
Ref.~\cite{Schmidt:2012az}. This evolution is used to dynamically populate an
interpolation grid which is used thereafter for performance reasons.

\paragraph{The analytic approximation} is given by the following expression, again up to
four-loop order:
\begin{multline}
 \alphaS(\mu) = \frac{1}{b_0 t} \biggl( 1 - \frac{b_1 \ln t}{b_0^2 t} + \frac{b_1^2 (\ln^2 t - \ln t - 1) + b_0 b_2}{b_0^4 t^2} - \\ % \linebreak[1]
  \frac{b_1^3 (\ln^3 t - \frac{5}{2} \ln^2 t - 2 \ln t + \frac{1}{2}) + 3 b_0 b_1 b_2 \ln t - 1/2 b_0^2 b_3}{b_0^6 t^3} \biggr),
\end{multline}
where $t = \ln\left(\mu^2/\LambdaQCD^2\right)$. Here \LambdaQCD takes distinct
values for different $n_f$, and these are required input parameters for the
number of active flavours that are desired in the calculation. General flavour
thresholds are possible with the analytic solver.

\paragraph{The interpolation option} uses a set of \alphaS values and their
corresponding $Q$ knots, provided as metadata, to interpolate using a log-cubic
interpolation with constant extrapolation for $Q^2 > Q^2_\textrm{last}$ and
logarithmic gradient extrapolation for $Q^2 < Q^2_\textrm{first}$. Discontinuous
subgrids are supported, to allow improved treatment of the impact of flavour
thresholds on \alphaS evolution.

These \alphaS evolution options are specified, cf.~the grid interpolators and
extrapolators, via an \kbd{AlphaS\_Type} metadata key on the PDF member or set. By
default the general PDF quark mass, $M_Z$, etc. metadata parameters are used for
\alphaS evaluation, but specific \kbd{AlphaS\_*} variants are also provided and
take precedence. Other details of the \alphaS scheme, such as variable or fixed
flavour number scheme, are specified by the \kbd{AlphaS\_FlavorScheme} and
\kbd{AlphaS\_NumFlavors}\footnote{Note that American spelling is used
  consistently in the LHAPDF\,6 interface.} keys. Quark thresholds can be
treated separately from the quark masses, but the latter are used as the default
thresholds.

%% file: usage.tex
%auto-ignore

In this section we give brief demonstrations of how to acquire and use PDF
objects in the three languages supported by LHAPDF\,6: C++, Python, and Fortran
(the latter via a legacy compatibility layer which provides the LHAPDF\,5
Fortran API, as will be described in Section~\ref{sec:compatibility}).

\subsection{C++ example}
\noindent
\kbd{\#include "LHAPDF/LHAPDF.h"}\\
\kbd{...}\\
\kbd{LHAPDF::PDF* p = LHAPDF::mkPDF("CT10/0");}\\
\kbd{cout <{}< p->xfxQ(21, 1e-4, 100.);}\\
\kbd{delete p;}\\
\kbd{vector<unique\_ptr<LHAPDF::PDF*>{}> ps \textbackslash}\\
\hspace*{2em}\kbd{= LHAPDF::mkPDFs("CT10nlo");}

\subsection{Python example}
\noindent
\kbd{import lhapdf}\\
\kbd{p = lhapdf.mkPDF("MSTW2008nlo68cl/1")}\\
\kbd{xfs = [p.xfxQ(pid, 1e-3, 100) for pid in p.flavors()]}\\
\kbd{s = lhapdf.getPDFSet("CT10nlo")}\\
\kbd{ps = s.mkPDFs()}

\subsection{Fortran example (same as for LHAPDF\,5)}
\noindent
\kbd{double precision x, q, f(-6:6)}\\
\kbd{x = 1.0D-4}\\
\kbd{q = 50.0D0}\\
% character name*64
% name = "CT10.LHgrid"
\kbd{call InitPDFsetByName("CT10.LHgrid")}\\
\kbd{call InitPDF(0)}\\
\kbd{call evolvePDF(x,Q,f)}

%% file: data.tex
%auto-ignore

LHAPDF\,6 uses a single system of metadata for all PDF sets, and a unified
interpolation grid format for all PDFs implemented via the \kbd{GridPDF} class
-- this is the case for all currently active PDFs, both all those migrated from
LHAPDF\,5 and the several new sets supplied directly to LHAPDF\,6.

All these data files, and an index file used to look up PDF members by a unique
global integer code -- the LHAPDF ID, following the scheme started by PDFLIB --
are searched for in paths which may be set via the code interface, which
falls back to the \env{LHAPDF\_DATA\_PATH} environment variable if set, then to
the legacy \env{LHAPATH} variable if set, and finally to the build-time
\var{install-prefix}\-\kbd{/share/LHAPDF/} data directory. The search paths set
via the API and via the environment variables may contain several different
locations, separated in the usual way by colon (\kbd{:}) characters in the
variables; as usual these are searched in left-to-right order, returning as soon
as a match is found.

Since it is shared between all prospective PDF implementations and can influence
the interpretation of the PDF data formats, we first describe the metadata
format in some detail, then the data format for\linebreak[20] LHAPDF\,6's standard
interpolation grids.

\subsection{Metadata format}

Metadata is encoded in LHAPDF\,6 using the standard YAML\,\cite{yamlweb} syntax,
and a uniform system is used for controlling system behaviours and storing PDF
physical information. YAML %(``Yet Another Markup Language'')
is a simple data structure syntax designed as a % less verbose and
more human readable/writeable variant of XML. % or JSON.
Its use in LHAPDF\,6 consists of dictionaries of key--value pairs, written as
\kbd{Key: Value}. The LHAPDF keys are all character strings; the value types may
be booleans, strings, integers, floating point numbers, or lists of numbers
written as \kbd{[1,2,3...]}. Valid boolean values include true, false, yes, no,
1, 0, and capitalised variants. The \kbd{yaml-cpp} package~\cite{yamlcppweb} is
embedded inside the LHAPDF library\footnote{With a modified namespace to avoid
  clashes with external usage.} and is responsible for parsing of the YAML data
sections, which are then available in C++ typed fashion from the \kbd{Info}
class and its specialisations.

Each PDF has a data file, the first part of which is YAML;
these files share a set directory with a \kbd{\var{setname}.info} file which is
in the same format; and lastly the global configuration lives in a
\kbd{lhapdf.conf} file, again in YAML.

As already mentioned, metadata keys set at a more specific level will override
those set more globally; it can hence be most efficient (for maintenance) to set
a not-quite ubiquitous key at \kbd{PDFSet} level and override it in the minority
of \kbd{PDF} members to which it does not apply. Major metadata keys and their
types are listed in Table~\ref{tab:metakeys}.

\begin{table*}[tp]
  \centering
  \caption{Main metadata keys used in LHAPDF\,6 along with their data types and descriptions. Full
    information on the standard metadata keys and their usage is found in the \kbd{CONFIGFLAGS}
    file in the LHAPDF code distribution, and on the LHAPDF website.}
  \label{tab:metakeys}
  \begin{tabular}{llll}
    \toprule
    Name & Type & Default value & Description\\
    \midrule
    \multicolumn{4}{l}{\textsc{Usually system-level}}\\
    \kbd{Verbosity}     & int & 1 & Level of information/debug printouts \\
    \kbd{Pythia6LambdaV5Compat} & bool & true & Return incorrect \LambdaQCD values in the PYTHIA6 interface \\
    \addlinespace
    \multicolumn{4}{l}{\textsc{Usually set-level}}\\
    \kbd{SetDesc} & str & & Human-readable short description of the PDF set \\
    \kbd{SetIndex} & int & & Global LHAPDF/PDFLIB PDF set ID code of first member \\
    \kbd{Authors} & str & & Authorship of this PDF set \\
    \kbd{Reference} & str & & Paper reference(s) describing the fitting of this PDF set \\
    \kbd{DataVersion} & int & -1 & Version number of this data, to detect \& update old versions \\
    \kbd{NumMembers} & int & & Number of members in the set, including central (0) \\
    \kbd{Particle} & int & 2212 & PDG ID code of the represented composite particle  \\
    \kbd{Flavors} & list[int] & & List of PDG ID codes of constituent partons in this PDF \\
    \kbd{OrderQCD} & int & & Number of QCD loops in calculation of PDF evolution \\
    \kbd{FlavorScheme} & str & & Scheme for treatment of heavy flavour (\kbd{fixed}/\kbd{variable})\\
    \kbd{NumFlavors} & int & & Maximum number of active flavours \\
    \kbd{MZ}            & real & 91.1876 & $Z$ boson mass in \GeV \\
    \kbd{MUp}, \dots, \kbd{MBottom}, \kbd{MTop} & real & 0.002, \dots, 4.19, 172.9 & Quark masses in \GeV \\
    \kbd{Interpolator}  & str & logcubic & Factory argument for interpolator making \\
    \kbd{Extrapolator}  & str & continuation & Factory argument for extrapolator making \\
    \kbd{ForcePositive} & int & 0 & Allow negative (0), zero (1), or only positive (2) $xf$ values \\
    \kbd{ErrorType} & str & & Type of error set (hessian/symmhessian/replicas/unknown)\\
    \kbd{ErrorConfLevel} & real & 68.268949\ldots & Confidence level of error set, in percent \\
    \kbd{XMin, XMax} & real & & Boundaries of PDF set validity in $x$ \\
    \kbd{QMin, QMax} & real & & Boundaries of PDF set validity in $Q$ \\
    \kbd{AlphaS\_Type}  & str & analytic & Factory argument for \alphaS calculator making \\
    \kbd{AlphaS\_MZ} & real & 91.1876 & $Z$ boson mass in \GeV, for $\alphaS(M_Z)$ treatment \\
    \kbd{AlphaS\_OrderQCD} & int & & Number of QCD loops in calculation of \alphaS evolution \\
    \kbd{AlphaS\_Qs,\_Vals} & list[real] & & Lists of $Q$ \& \alphaS interpolation knots\\
    \kbd{AlphaS\_Lambda4/5} & real & & Values of $\LambdaQCD^{(4)}$ and $\LambdaQCD^{(5)}$ for analytic \alphaS\\
    \addlinespace
    \multicolumn{4}{l}{\textsc{Usually member-level}}\\
    \kbd{PdfType} & str & & Type of PDF member (\kbd{central}/\kbd{error}/\kbd{replica}) \\
    \kbd{Format}  & str & & Format of data grid (\kbd{lhagrid1}/\kbd{\dots}) \\
    \bottomrule
  \end{tabular}
\end{table*}

\subsubsection{System-level metadata}

The basic system-level configuration is set by a collection of metadata keys in
the file \kbd{lhapdf.conf} -- specifically the first file of that name to be
found in the runtime search path, as is the case for all file lookup in
LHAPDF\,6. The system-level metadata can be obtained by loading the
generic \kbd{info} object using the\linebreak[20] \kbd{LHAPDF::getConfig()} function.

The default set of such keys is relatively small and sets some uncontroversial
values such as use of the log-cubic interpolator and the continuation
extrapolator, and default quark and $Z$ boson masses.

The \kbd{Verbosity} key is also set here: this integer-valued parameter controls
the level of output written to the terminal on loading PDFs and performing other
operations, and by default is set to 1, which produces a small announcement on
first loading a PDF set; by comparison 0 is silent and 2 produces more detailed
and more frequent print-outs.

\subsubsection{Set-level metadata}

As opposed to LHAPDF\,5, where each PDF set was encoded in a single text data
file, the LHAPDF\,6 format is that each set is a \emph{directory} with the same
name as the set, which contains one \kbd{\var{setname}.info} file, plus the
member-specific data files. The common set-level metadata should be set in the
\kbd{.info} file. The set-level metadata can be obtained by loading the
lightweight \kbd{PDFSet} object using the \kbd{LHAPDF::getPDFSet()} function.

The bulk of metadata should be declared at the PDF set level, except in those
sets where each member has a systematic variation in the information set via
metadata keys such as quark masses/thresholds and \alphaS. The information
typically specified at the set-level includes quark and $Z$ masses (even if the
system-level defaults are appropriate, it is safest to repeat the values used
for future-proofing), the PDG ID code of the parent particle (to allow for
identifiable nuclear PDFs in future), and the error treatment, confidence level,
etc. of the systematic uncertainty variations in the set, to permit automated
error computation such as that described in Section~\ref{sec:uncertainties}.

\subsubsection{Member-level metadata}

As will be described in more detail below, in addition to the \kbd{.info} file
in each PDF set directory, there is one ``.dat'' file for each PDF member in the
set. This structure permits much faster lookup of set-level metadata and random
access to single members in the set, compared to the one-file-per-set structure
used by LHAPDF\,5.

The top section of each \kbd{.dat} file is devoted to member-level metadata in
the usual format. This should contain the \kbd{Format} metadata key which will
be used to determine what sort of PDF is being loaded and trigger the
appropriate constructor (e.g. \kbd{GridPDF}, for key value \kbd{lhagrid1}) via a
factory function to read the rest of the file. This header section ends with a
mandatory line containing only three dash characters (\kbd{-{}-{}-}), the
standard YAML sub-document separator. The \kbd{PdfType} key is also usually set
here, to declare whether this member is a central or error/replica PDF
member. Any other metadata key may also be declared at member-level, possibly
overriding set-level values; this is particularly the case for special quark
mass or \alphaS systematic variation sets.

PDF member-level metadata can be loaded without needing to load the much larger
data block by use of the \kbd{LHAPDF::mkPDFInfo(...)} factory functions.

\subsection{PDF grid data format}

Within the \kbd{\var{setname}} directory, each PDF member has its own file named
\kbd{\var{setname}{}\_\var{nnnn}.dat}, where \var{nnnn} is a 4-digit zero-padded
representation of the member number within the set -- for example member 0 is
``0000'' and member 51 is ``0051'' -- reasonably assuming that there will be no
need for PDF sets with more than 10,000 members. The ``central'' PDF set member
must always be number 0.

The splitting of PDF set data into one file per member permits faster random
access to individual members (the central member being the most common), and
permits an extreme space optimisation for circumstances which require it: PDF
data directories may be cut down to only contain the subset of members which are
going to be used. While not generally recommended, this may give a significant
space saving and be useful for resource-constrained applications -- for example,
to allow LHC experiments' Grid installations to contain the central members of
many PDF sets where distribution of the full sets would make unreasonable
demands on Grid sites and kit distribution.

As already described, the first section in each \kbd{.dat} file contains a YAML
header of member-specific metadata, until the \kbd{-{}-{}-} separator
line. After this line, the grid data begins. Each subgrid in $Q$ is treated
separately and should be listed in the file in order of increasing $Q$ bin,
separated again by \kbd{-{}-{}-} separator lines. The file must be terminated by
such a line after the last subgrid data block.

Within each subgrid block there is a three-line header then a large number of
lines giving the PDF values at each $(x, Q)$ point. The first line in the header
is a space-separated, ordered list of $x$ knot values; the second is a similar list of
$Q$ knot values; and the third is a list of the particle ID codes to be given in
the data block to follow. Note that although the interpolator/extrapolator
implementations operate canonically in $Q^2$ (or $\log Q^2$) to avoid expensive
square-root function calls in typical usage, in the data files we always use $Q$
to give the scale: this is for ease of interpretation and debugging, since
physicists find it more natural to interpret scales related to e.g. the masses
or transverse momenta of produced particles than the squares of such
quantities. The particle codes listed on the third header line are in the
standard PDG ID scheme, and must be given in the order that columns of PDF
values will be presented in the remainder of the subgrid block. It is
anticipated that the ``generator specific'' range of PDG ID codes may be used in
future to permit valence/sea decompositions or aliasing of PDF components in the
LHAPDF data files, but there has not yet been demand for such features.

The gridded PDF value data comes next, with each line giving an $xf(x;Q)$ value
for each of the parton ID codes given in the final line of the block header.
The order of lines corresponds to a nested pair of loops over the $x$ and $Q$
knot lists given in the block header, e.g. what would result from the pseudocode\\[1ex]
\textsf{\hspace*{1em} for $x$ in $\{x\}$:\\ \hspace*{2em} for $Q$ in $\{Q\}$: \\
  \hspace*{3em} write $xf_i(x;Q^2)$ for $i$ in PIDs}\\[1ex]
The lines hence come in groups of lines with fixed $x$, each group containing as
many lines as there are $Q$ knots, with the total subgrid containing $|\{x\}|
\times |\{Q\}|$ lines of $xf$ grid data in addition to the three header lines
that specify the knot positions and parton flavours. The \kbd{GridPDF} parser
makes many consistency checks on the correctness of the format.

% The interpolation PDF data files use a single uniform plain text format. This
% starts with a YAML-format header section (to provide member-specific metadata
% overrides), then blocks of numbers for each $x$--$Q^2$ subgrid, each started by
% a header declaring the knot positions and encoded parton flavours.
%  If desired, the member content of a set
% could be stripped down to save space in special applications.

\subsection{\alphaS interpolation data format}

If the interpolation scheme is used for getting \alphaS values from a PDF
(\kbd{AlphaS\_Type} = \kbd{ipol}), the interpolation knot \alphaS values and $Q$
positions are given as lists of floating point values for the metadata keys
\kbd{AlphaS\_Vals} and \kbd{AlphaS\_Qs} respectively. These are used for
log-cubic interpolation in the usual way. Naturally the two lists must be of the
same length. Subgrid boundaries in $Q$ are expressed by a repetition of the
boundary $Q$ value -- the corresponding \alphaS values should be given as the
\alphaS limits from below and above the boundary.

\subsection{Index file}

The \kbd{LHAPDF::mkPDF(int)}, \kbd{LHAPDF::lookupPDF(int)},
and\linebreak[20] %\penalty-100
\kbd{LHAPDF::lookupLHAPDFID(string, int)} factory functions\linebreak[20] make
use of the global LHAPDF ID code and its mapping to PDF members. This mapping is
done via the \kbd{pdfsets.index} file, which must be found in the search paths
for these lookup functions to work. This file contains three data columns
separated by whitespace: the LHAPDF ID, the set name, and the set's latest data
version. The only entries in the index file are the first entries in each PDF
set, since the ID codes and containing sets of any member may be extracted from
these.

The LHAPDF ID index codes are given in each PDF set \kbd{.info} file via the
\kbd{SetIndex} metadata key, which gives the LHAPDF ID number of the first
(central) member in the set. To ease maintenance work and minimise errors, the
index file is generated automatically by loading and querying the \kbd{.info}
files from all the PDF sets. LHAPDF's online documentation of available PDF sets
is also generated by this method.

\subsection{Distribution and updating}

LHAPDF\,6 breaks the tight binding of PDF data files and the LHAPDF code
library: releases of new PDF set data now happens in general out of phase with
software releases, permitting much faster release of PDF sets for use via
LHAPDF. This was a major design goal of LHAPDF\,6.

The sets are distributed as \kbd{\var{setname}.tar.gz} archive files, each one
expanding to the \kbd{\var{setname}} directory which contains the set's metadata
(\kbd{.info}) and data (\kbd{.dat}) files. A typical PDF set with 50 members and
5 quark flavours corresponds to a 5--10\,MB compressed tarball, which on
expansion will occupy 20--30\,MB. The 100-member NNPDF sets, which also include
top (anti)quark PDFs, are somewhat larger at $\mathcal{O}(30\,\text{MB})$
compressed and $\mathcal{O}(80\,\text{MB})$ expanded; sets with fewer members or
fewer flavours require correspondingly less disk space. Directly using the
unexpanded tarballs is not supported, but a trick to do so will be described in
Section~\ref{sec:performance}.

% As there is no set-specific or family-specific handling code in the library,
% new PDFs may be made and privately tested (or even released separately from
% LHAPDF) without needing a new version of the LHAPDF library.
The only update required for full usability of a new PDF set is an updated
version of the \kbd{pdfsets.index} file, although this is only needed for PDF
use via the LHAPDF ID code: access to PDFs by set-name + set-member number does
not use the index file and is encouraged for robustness and human readability.
New official PDF set data will be uploaded to the LHAPDF
website\,\cite{lhapdfweb} along with an updated, automatically generated version
of the \kbd{pdfsets.index} file. Official PDF sets will also be distributed,
both tarballed and expanded, via the CERN AFS and CVMFS distributed file
systems.

Officially supported PDF sets must contain the\linebreak[20] \kbd{DataVersion}
integer metadata key to allow for tracking of bugfix releases of the set data
files. The latest such number is written into the \kbd{pdfsets.index} file, and
can be used to detect when an update is available for a PDF set installed on a
user's system. LHAPDF\,6 provides and installs a PDF data management script
simply called \kbd{lhapdf}, with an interface similar to the Debian/Ubuntu Linux
\kbd{apt-get} command: calling \kbd{lhapdf list} and \kbd{lhapdf install} will
respectively list and install PDFs from the Web, \kbd{lhapdf update} will
download the latest index file from the LHAPDF website, and \kbd{lhapdf upgrade}
will download updated versions of PDF set files if notified as available in the
current index file. The rest of the script features are interactively documented
by calling \kbd{lhapdf -{}-{}help}.

In future PDF sets may be released which require LHAPDF features such as newer
grid formats, which are only available after a particular LHAPDF release. In
this situation, which has not yet been encountered, the set should declare the
\kbd{MinLHAPDFVersion} metadata flag to have an integer value corresponding to
the earliest LHAPDF\,6 version with which it is compatible. This integer version
code will be described in Section~\ref{sec:compatibility}.

%% file: uncertainties.tex
%auto-ignore

Over the last decade or so, it has become standard practice for PDF fits to
propagate the experimental uncertainties on the fitted data points and provide a
number of alternative PDF members in addition to the central member.  An
estimate of PDF uncertainties on either the PDFs themselves, or derived
quantities such as parton luminosities or cross-sections, can then easily be
calculated with a simple formula using the quantity calculated for all members
of the PDF set.  Correlations between two quantities can also be calculated, for
example, to establish the sensitivity of a particular cross-section to a PDF of
a particular flavour.  However, in practice, there are multiple formulae in
common use depending on the PDF set together with a variety of different
confidence levels, requiring some specialist knowledge from the user in order to
apply the correct formula, and potentially leading to mistakes by non-experts
that could severely underestimate or overestimate the importance of PDF
uncertainties.  Moreover, each user or code that calculates PDF uncertainties
needs to implement the correct formula for each PDF set and possibly rescale
uncertainties to a desired confidence level, typically with branching based on
the name of the PDF set, resulting in a vast duplication of effort.

Starting from LHAPDF 5.8.8 first steps were taken towards a more automatic
calculation of PDF uncertainties by providing Fortran subroutines
\kbd{GetPDFUnc\-Type}, \kbd{GetPDFuncertainty} and
\kbd{GetPDFcorrelation} that would attempt to use the appropriate formulae
based on the name of the grid format.  However, C++ versions of these functions
were not implemented and it was not straightforward to discern the confidence
level of a given PDF set.  Starting from LHAPDF 6.1.0 member functions were
implemented in the \kbd{PDFSet} class making use of the new set-level
metadata, specifically \kbd{ErrorType} and \kbd{ErrorConfLevel}, with
several extensions to the original Fortran subroutines.  Here we describe these
functions and the formulae implemented based on the chosen PDF \kbd{set}, for
each of the three currently supported values of \kbd{ErrorType}, namely
\kbd{hessian}, \kbd{symm\-hessian} or \kbd{replicas}.\footnote{The more
complicated prescription for the HERAPDF/ATLAS ``\kbd{VAR}'' model and
parametrisation errors differs between the different sets and is not
currently supported.}  An example
program (\kbd{testpdf\-unc.cc}) demonstrates the basic functionality.  See,
for example, Section 2.2.3 of Ref.\,\cite{Forte:2013wc} for a more comprehensive
review of the different approaches, and Refs.\,\cite{Watt:2011kp,Watt:2012tq}
for more discussion of the relevant formulae.

\subsection{\kbd{set.uncertainty(values, cl, alternative)}}

This function takes as input a vector of \kbd{values} and returns a
\kbd{PDFUncertainty} structure containing a \kbd{central} value,
asymmetric (\kbd{errplus} and \kbd{errminus}) and symmetric
(\kbd{errsymm}) uncertainties, and the \kbd{scale} factor used to rescale
uncertainties to the desired confidence level (\kbd{cl}, in percent), by
default $1\text{-sigma} = \mathrm{erf}(1/\sqrt{2}) \simeq 68.268949\%$.  The
formulae used for the calculation depend on the value of \kbd{ErrorType} and
are hidden from the user, but for reference we give the different formulae below
for each \kbd{ErrorType}.  The \kbd{alternative} option is only relevant
for the \kbd{replicas} case.
\begin{description}
\item[\kbd{hessian} : ] Given a central PDF member $S_0$ and $2N_{\rm par}$
  eigenvector PDF members $S_i^\pm$ ($i=1,\ldots,N_{\rm par}$), where $N_{\rm
    par}$ is the number of fitted parameters, the central value $F_0$ and
  asymmetric ($\sigma_F^\pm$) or symmetric ($\sigma_F$) PDF uncertainties on a
  PDF-dependent quantity $F(S)$ are given by:
  \begin{align}
    F_0 &= F(S_0),\quad F_i^+ = F(S_i^+),\quad F_i^- = F(S_i^-),\\
    \sigma_F^+ &= \sqrt{\sum_{i=1}^{N_{\rm par}} \left[{\rm max}\left(\;F_i^+-F_0,\;F_i^--F_0,\;0\right)\right]^2},\\
    \sigma_F^- &= \sqrt{\sum_{i=1}^{N_{\rm par}} \left[{\rm max}\left(\;F_0-F_i^+,\;F_0-F_i^-,\;0\right)\right]^2},\\
    \sigma_F &= \frac{1}{2}\sqrt{\sum_{i=1}^{N_{\rm par}} \left(F_i^+-F_i^-\right)^2}. \label{eq:symmunc}
  \end{align}

\item[\kbd{symmhessian} : ] For the simpler case where only a central PDF
  member $S_0$ and $N_{\rm par}$ eigenvector PDF members $S_i$
  ($i=1,\ldots,N_{\rm par}$) are provided, the central value and PDF
  uncertainties are calculated as:
  \begin{align}
    F_0 &= F(S_0),\quad F_i = F(S_i),\\ \sigma_F^+ = \sigma_F^- = \sigma_F &= \sqrt{\sum_{i=1}^{N_{\rm par}} \left(F_i-F_0\right)^2}.
  \end{align}
\item[\kbd{replicas} : ] Given a set of $N_{\rm rep}$ equiprobable Monte
  Carlo replica PDF members $\mathcal{S}^k$ ($k=1,\ldots,N_{\rm rep}$), created
  either by making fits to randomly shifted data points or by randomly sampling
  the parameter space, the central value and PDF uncertainties are by default
  (\kbd{alternative=false}) given by the average and standard deviation over
  the replica sample:
  \begin{align}
    F_0 & = \langle F\rangle = \frac{1}{N_{\rm rep}}\sum_{k=1}^{N_{\rm rep}}F(\mathcal{S}^k), \label{eq:MCav}\\
    \sigma_F^+ = \sigma_F^- = \sigma_F & = \sqrt{\frac{1}{N_{\rm rep}-1}\sum_{k=1}^{N_{\rm rep}}\left[F(\mathcal{S}^k)-F_0\right]^2} \nonumber \\
    & = \sqrt{\frac{N_{\rm rep}}{N_{\rm rep}-1}\left[\langle F^2\rangle-\langle F\rangle^2\right]}. \label{eq:MCsd}
  \end{align}
  Alternatively (if \kbd{alternative=true}), a confidence interval (with
  level \kbd{cl}) is constructed from the probability distribution of
  replicas, with the central value $F_0$ given by the median, then the interval
  $[F_0-\sigma_F^-,F_0+\sigma_F^+]$ contains \kbd{cl}\% of replicas, while
  the symmetric uncertainty is simply defined as
  $\sigma_F=(\sigma_F^++\sigma_F^-)/2$.
\end{description}
% TO DO: possibly make some plots e.g. of gluon distribution, showing individual members and uncertainties, for Hessian case (e.g. CT10) and replicas case (e.g. NNPDF3.0) with both "alternative" options.

\subsection{\kbd{set.correlation(valuesA, valuesB)}}

This function takes as input two vectors \kbd{valuesA} and \kbd{valuesB},
containing values for two quantities $A$ and $B$ computed using all PDF members,
and returns the correlation cosine $\cos\phi_{AB}\in[-1,1]$.  Values of
$\cos\phi_{AB}\approx 1$ mean that $A$ and $B$ are highly correlated, values of
$\approx -1$ mean that they are highly anticorrelated, while values of
$\approx 0$ mean that they are uncorrelated.  Again, we give the different
formulae below for each \kbd{ErrorType}, although these formulae are
invisible to the user.

\begin{description}
\item[\kbd{hessian} : ] The correlation cosine is calculated as:
  \begin{equation}
    \cos\phi_{AB} = \frac{1}{4\,\sigma_A\,\sigma_B}\,\sum_{i=1}^{N_{\rm par}}\,\left(A_i^+-A_i^-\right)\,\left(B_i^+-B_i^-\right),
  \end{equation}
  where the uncertainties $\sigma_A$ and $\sigma_B$ are calculated using the symmetric formula, Eq.~\eqref{eq:symmunc}.
\item[\kbd{symmhessian} : ]
  Similarly, the correlation cosine is:
  \begin{equation}
    \cos\phi_{AB} = \frac{1}{\sigma_A\,\sigma_B}\,\sum_{i=1}^{N_{\rm par}}\,\left(A_i-A_0\right)\,\left(B_i-B_0\right).
  \end{equation}
\item[\kbd{replicas} : ]
  In the Monte Carlo approach:
  \begin{equation}
    \cos\phi_{AB} = \frac{N_{\rm rep}}{N_{\rm rep}-1}\frac{\langle AB\rangle-\langle A\rangle\langle B\rangle}{\sigma_A\,\sigma_B},
  \end{equation}
  where the average $\langle A\rangle$ and standard deviation $\sigma_A$ are defined in Eqs.~\eqref{eq:MCav} and \eqref{eq:MCsd}, respectively.
\end{description}

\subsection{\kbd{set.randomValueFromHessian(values, randoms, symmetrise)}}
This function will generate a random value from a vector of \kbd{values},
containing values for a quantity $F$ computed using all PDF members of a
\kbd{hessian} (or \kbd{symm\-hessian}) PDF set, and a vector of random
numbers \kbd{randoms} sampled from a Gaussian distribution with mean zero and
variance one.  Random values generated in this way\,\cite{Watt:2012tq} can
subsequently be used for applications such as Bayesian
reweighting\,\cite{Ball:2010gb,Ball:2011gg,Paukkunen:2014zia}
or combining predictions from different PDF fitting groups
(as an alternative to taking the envelope)\,\cite{Forte:2013wc}.  Below
we give the formulae used for each relevant \kbd{ErrorType}.
\begin{description}
\item[\kbd{hessian} : ] For the option \kbd{symmetrise=false}, we build a
  random value of a quantity $F$ according to:
  \begin{equation}
    F^k = F(S_0) + \sum_{j=1}^{N_{\rm par}}\left[F(S_j^\pm)-F(S_0)\right]\,|R_j^k|,
  \end{equation}
  where either $S_j^+$ or $S_j^-$ is chosen depending on the sign of the Gaussian
  random number $R_j^k$.  We can repeat this procedure to generate $N_{\rm rep}$
  random values, where $k=1,\ldots,N_{\rm rep}$.  However, this asymmetric
  prescription means that the average $\langle F\rangle$ over the $N_{\rm rep}$
  values does not tend to the best-fit $F(S_0)$ for large values of $N_{\rm rep}$.
  Hence the default behaviour (\kbd{symmetrise=true}) is to use a symmetrised
  formula ensuring this condition:\footnote{This formula corrects
    Eq.~(6.5) of Ref.~\cite{Watt:2012tq} to preserve correlations by not
    taking the absolute value of the quantity in square brackets.}
  \begin{equation}
    F^k = F(S_0) + \frac{1}{2}\sum_{j=1}^{N_{\rm par}}\left[F(S_j^+)-F(S_j^-)\right]\,R_j^k.
  \end{equation}

\item[\kbd{symmhessian} : ] In this case the \kbd{symmetrise} option has
  no effect and the formula is:
  \begin{equation}
    F^k = F(S_0) + \sum_{j=1}^{N_{\rm par}}\left[F(S_j)-F(S_0)\right]\,R_j^k.
  \end{equation}
\end{description}
An example program (\kbd{hessian2replicas.cc}) is provided that uses the
\kbd{randomValueFromHessian} function to convert an entire \kbd{hessian}
(or \kbd{symmhessian}) PDF set into a corresponding PDF set of Monte Carlo
\kbd{replicas}.

%% file: reweighting.tex
%auto-ignore

A common use of PDFs is reweighting of event samples to behave as if they had
originally been generated with PDFs other than the one that was actually
used. This is particularly an effective strategy when applying a PDF uncertainty
procedure such as the PDF4LHC recommendation~\cite{Botje:2011sn} which involves
predictions from $\sim 200$ PDF members -- generating 200 independent MC samples
is unrealistic and hence reweighting is a common approach. The reweighting
factor for a leading-order hadron--hadron process from PDF $xf(x;Q^2)$ to PDF
$xg(x;Q^2)$ is defined as
\begin{equation}
  \label{eq:reweight}
  w =
  \frac{x_1 g_{i/B1}(x_1; Q^2)}{x_1 f_{i/B1}(x_1; Q^2)} \cdot
  \frac{x_2 g_{j/B2}(x_2; Q^2)}{x_2 f_{j/B2}(x_2; Q^2)} .
\end{equation}

But we must note limitations in this strategy: a single well-defined set of
partonic initial conditions is only defined at tree level, where there are no
real- and virtual-emission counter-terms to deal with. Reweighting higher-order
calculations where counter-terms are involved requires deeper knowledge of the
event generation than is typically available to users who wish to
retrospectively reweight an existing event sample -- it is much more
appropriately done by the NLO MC generator code itself, and this is supported by
at least the Sherpa\,\cite{Gleisberg:2008ta}, POWHEG-BOX\,\cite{Alioli:2010xd}, and
MadGraph5\_\-aMC@NLO\,\cite{Alwall:2014hca} generator packages.

Further limitations are that PDF reweighting is typically applied only at the
fixed-order matrix element level. Parton-shower-matched event simulations also
include PDF terms in the Sudakov form factors that appear in initial-state
radiation emission probabilities, and these should strictly also be reweighted
-- but doing so consistently would require a sum over possible emission
histories, which has yet to be formalised or implemented in such programs. And
finally there is the issue of \alphaS consistency: if reweighting PDFs then
appearances of the strong coupling -- ideally both in the matrix element and
shower -- should also be reweighted. As this tends not to be done, PDF
reweighting should only be done between PDFs with similar \alphaS values in the
scale range of the process. In particular reweightings between LO and NLO PDFs,
which tend to have very different \alphaS values, are strongly discouraged.

LHAPDF\,5 provided no built-in support for reweighting, since the operation in
Eq.~\eqref{eq:reweight} is numerically trivial. However it has transpired that
within experimental collaborations there was demand for a ``tool'' to assist
with this calculation. In the interests of usability LHAPDF\,6 hence provides
helper functions for computation of reweighting factors, in the
\kbd{LHAPDF/Reweighting.h} header file. These are divided into two categories --
single-beam functions which calculate the individual weighting factors for each
beam, and two-beam functions which multiply together the weights for the two
beams. The single-beam function signature is \kbd{LHAPDF::weightxQ2(i, x, Q2,
  pdf\_f, pdf\_g, aschk=0.05)}, which will reweight\linebreak[100]
$x f_i(x; Q^2) \to x g_i(x; Q^2)$. The optional \kbd{aschk} argument gives a
threshold for the relative difference in $\alphaS(Q^2)$ between the two PDFs
before the LHAPDF system will print a warning: this tolerance may be set
negative to disable checking, but this is not advised for physics reasons. The
\kbd{pdf\_f,g} arguments to this function may be given either as (\kbd{const})
references to \kbd{PDF} objects or as any kind of (smart or raw) \kbd{PDF}
pointer.  The equivalent two-beam functions have the same form, only generalised
to have two parton ID and two $x$ arguments.

%% file: compatibility.tex
%auto-ignore

Due to the ubiquity of LHAPDF as a source of PDF information in HEP software, it
would be unrealistic to release LHAPDF\,6 without also providing a route for
this mass of pre-existing code to continue to work.

\subsection{Legacy code interfaces}
To this end, legacy interfaces have been provided to the Fortran LHAPDF and
PDFLIB interfaces, and to the LHAPDF\,5 C++ interface. These are written in C++,
and following the naming used in LHAPDF\,5 to denote the backward compatibility
interface with PDFLIB, are called the ``LHAGlue'' interface. It is entirely
localised to the \kbd{LHAGlue.h} and \kbd{LHAGlue.cc} files within LHAPDF\,6.

The Fortran compatibility interfaces are implemented in C++ using \kbd{extern
  "C"} linkage and the GCC Fortran symbol mangling conventions. Since there is a
mismatch between the unlimited, dynamic memory allocation model of LHAPDF\,6's
native C++ interface and the static, pre-allocated slots model of LHAPDF\,5, a
state machine was implemented to manage \kbd{PDF} object creation and deletion
in numbered slots via the Fortran LHAPDF\,5 \kbd{initpdfsetm} and \kbd{initpdfm}
routines. For simplicity many of the C++ LHAPDF\,5 API functions were
implemented via calls to these Fortran state-machine functions to reproduce the
LHAPDF\,5 behaviour.

Since the data format has changed in LHAPDF\,6 and there are no longer any data
files with the LHAPDF\,5 \kbd{.LHpdf} or \kbd{.LHgrid} file extensions, calls to
\kbd{initpdfsetm} which specify a name with such an extension will simply have
it stripped off before continuing with PDF loading. There is a special case of
this for the CTEQ6L1 PDF\,\cite{Pumplin:2002vw}, which was accidentally
implemented in\linebreak[20] LHAPDF\,5 with the mis-spelt name
\kbd{cteq6ll.LHpdf}: this name will automatically be translated to the correct
name, \kbd{cteq6l1}, by which it is called in LHAPDF\,6.

The legacy interfaces also contain a special case behaviour in the reporting of
$\LambdaQCD^{(4)}$ and $\LambdaQCD^{(5)}$, which never worked correctly for the
LHAPDF\,5 PDFLIB-type common-block interface to
PYTHIA\,6\,\cite{Sjostrand:2006za}. This value reporting is fixed in LHAPDF\,6,
but in the meantime many tunes of PYTHIA\,6's physics modelling have been built
around the assumption that an invalid value would be reported and PYTHIA would
default to 0.192, the $\LambdaQCD^{(4)}$ value of the CTEQ5L
PDF\,\cite{Lai:1999wy}. Since\linebreak[20] PYTHIA\,6 is itself now largely replaced by its
successor, Pythia\,8\,\cite{Sjostrand:2007gs}, and it is important that many of
these tunes continue to work with an implicitly incorrect \LambdaQCD value, a
boolean metadata key \kbd{Pythia6LambdaV5Compat} has been provided to trigger
the old physically incorrect but practically convenient behaviour. This flag is
set true by default in the system \kbd{lhapdf.conf} file, and may be changed in
this file or by runtime use of the metadata API.

\subsection{Version detection hooks}
As well as these compatibility interfaces, LHAPDF\,6 provides mechanisms to
allow C++ applications which use LHAPDF\,5 to detect which version they are
compiling against and hence migrate smoothly to the new version. Three C++
preprocessor macros are provided for this purpose:
\begin{description}
\item[\kbd{LHAPDF\_VERSION}] provides a string version of the 3-integer release
  version tuple (cf. the current release 6.1.4);
\item[\kbd{LHAPDF\_VERSION\_CODE}] is a version of this information encoded into
  a single integer by multiplying the first and second numbers by 10000 and 100
  respectively, then adding the three numbers together (making the 6.1.4 release
  have a single-integer code of 60104);
\item[\kbd{LHAPDF\_MAJOR\_VERSION}] is the first number in the version 3-tuple,
  as an integer (i.e. 6 for version 6.1.4).
\end{description}
These macros can be portably accessed by \kbd{\#include}'ing the
\kbd{LHAPDF/LHAPDF.h} header, which is available in both version 5 and version
6, and the integer codes can be used as a preprocessor test to separate code for
calling LHAPDF\,5 routines from the new, more powerful LHAPDF\,6 ones, for
example:\\[1ex]
\kbd{\#include "LHAPDF/LHAPDF.h"}\\
\kbd{\#if LHAPDF\_MAJOR\_VERSION == 6}\\
\hspace*{1em}\textit{\var{LHAPDF\,6 code}}\\
\kbd{\#else}\\
\hspace*{1em}\textit{\var{LHAPDF\,5 code}}\\
\kbd{\#endif}

\subsection{Uptake and prospects}
The legacy interfaces have been successfully tested with a variety of
widely-used MC generator codes, including PYTHIA\,6\,\cite{Sjostrand:2006za},
HERWIG\,6\,\cite{Corcella:2000bw},
POWHEG-BOX\,\cite{Alioli:2010xd},\linebreak[20] and
aMC@NLO\,\cite{Alwall:2014hca}. The main C++ parton shower generators, from
Sherpa~2.0.0\,\cite{Gleisberg:2008ta}, Herwig++~2.7.1\,\cite{Bahr:2008pv}, and
Pythia~8.200\,\cite{Sjostrand:2007gs} onwards all support LHAPDF\,6 via the
native C++ API. The global LHAPDF~ID code is still in use and will continue to
be allocated for submitted PDFs, meaning that the PDFLIB and LHAPDF\,5 Fortran
interfaces can continue to be used for some time, and will now return more
correct values in some circumstances (e.g. \alphaS values in multi-set
mode). % , and a consistent treatment of the number
% of set members to always include the central member).

% \TODO{Note fixed problems: Support for different set sizes. \alphaS etc. in
%   multi-set mode. $\Lambda$ reporting to PYTHIA and compat mode. Number of
%   members now reported correctly \& consistently.}

An improved Fortran interface to LHAPDF\,6 is intended but has not yet progressed
beyond initial stages; we welcome input from the Fortran MC generator community
in particular on what features they would like to see.

%% file: performance.tex
%auto-ignore

The re-engineering of LHAPDF has impact upon the memory and CPU performance of
the library. The main performance target in the redesign was to greatly reduce
the multiple-GB static memory requirement of an LHAPDF\,5 build with full
multiset functionality. We describe the effect on this performance metric in the
following section, and also mention the impact on CPU performance and data-file
disk space requirements. We also describe some possible avenues for further
performance improvements.

\subsection{Memory requirements}
The memory problems of LHAPDF\,5 fundamentally stem from the Fortran~77
limitation to static memory allocation, and the use of large static arrays for
PDF value interpolation in each PDF family's ``wrapper'' routine (i.e. the code
which interfaced the native PDF group code into the LHAPDF\,5 framework). By the
time of LHAPDF 5.9.1, the proliferation of such wrapper routines meant that
2.04\;GB of static memory was declared as required by the \kbd{libLHAPDF}
library. This static memory requirement was incompatible with LHC computing
systems, and the restricted memory builds used to work around process accounting
limits were suitable only for the most basic sort of event generation; working around
LHAPDF's technical limitations became a rite of passage in LHC data analysis.

% In practice this much RAM was not actually required to run LHAPDF on a typical
% workstation, since the large arrays were not initialised until first used and
% operating systems are typically intelligent enough to not allocate physical
% memory until assignment. But the static declaration of such large memory
% requirements proved disastrous for Grid usage, where process accountancy led to
% batch jobs with a full-featured LHAPDF\,5 being entirely blocked from
% execution. The interim solution was to build a very stripped-down,
% limited functionality version of LHAPDF, with the penalty that it was only
% suitable for the most basic sort of event generation. Working around LHAPDF's
% technical limitations became a standard rite of passage in LHC data analysis.

The dynamic memory model in LHAPDF\,6 completely solves this problem, as
illustrated by the static memory information obtained by running the \kbd{size}
command on the equivalent libraries between versions 5 and 6 of LHAPDF: this
information is shown in Table~\ref{tab:staticmem}. All static memory
requirements have been greatly reduced by the version 6 redesign, and the total
static memory footprint is now just 280\;kB, but the headline figure is the
reduction in static uninitialised data size from more than 2\;GB to a negligible
1.6\;kB. This does not reflect the total memory requirements of LHAPDF\,6 in
active use -- allocating a \kbd{GridPDF} will typically require a few hundred
kB, and loading a whole set into memory will require
$\mathcal{O}(10\;\text{MB})$, but the user is now fully in control of when they
allocate and deallocate that memory, as well as being \emph{able} to load single
PDF set members, an option not available in LHAPDF\,5.

\begin{table}[t]
  \centering
  \caption{Static memory requirements in kB for LHAPDF version 5 and 6 before any PDF allocation,
    broken down into the requirements for function, initialised data, and uninitialised
    data. LHAPDF\,6 is much lighter on all counts, but the overwhelmingly most important
    number is the reduction in uninitialised data from more than 2\;GB down to less than
    1\;MB. LHAPDF\,6 memory only becomes substantial when \kbd{PDF} objects are created,
    and is proportional to the grid sizes of those PDFs.}
  \label{tab:staticmem}
  \begin{tabular}{lrrr}
    \toprule
    Version & Functions & Init. data & Uninit. data\\
    \midrule
    5.9.1 & 1509.1 & 142.0 & \textbf{2039405.4}\\
    6.1.5 &  265.3 &   8.5 &       \textbf{1.6}\\
    \bottomrule
  \end{tabular}
\end{table}

\subsection{CPU performance}

LHAPDF\,6 was not specifically engineered for CPU performance gains, since this
was not typically a severe issue with LHAPDF\,5. However, particularly because
of the approach taken to multiple parton-flavour evolution in \kbd{GridPDF}
interpolation, there is some impact on CPU performance.

In LHAPDF\,5 the performance was dependent on which PDF set was being used, as
each wrapper routine was implemented independently and some were better
optimised than others; however, the \kbd{evolvePDF} and \kbd{xfx} routines
always returned a 13-element array of PDF values for the gluon + $2 \times 6$
quark flavours. They hence tended to be implemented such that the $x$--$Q^2$
``positional'' part of the interpolation weights was only computed once, rather
than being redundantly recomputed for every flavour at that point. This means
that LHAPDF\,6 interpolation is currently slightly slower than for LHAPDF\,5 if
all flavours are evaluated at every $(x,Q^2)$ point; however, if only one
flavour is required at a phase space point, then LHAPDF\,6 is significantly
faster since it does not need to interpolate an extra 12 values which will not
be used. Legacy code written to use the PDFLIB or LHAPDF\,5 interfaces is often
structured to make use of this feature, and such code may be slightly slower
with LHAPDF\,6. However, where code can be rewritten to make use of a
single-flavour approach, significant speed-ups can be achieved, as shown in
Table~\ref{tab:sherpacpu} which gives timing information obtained with the
Sherpa event generator\,\cite{Gleisberg:2008ta}. Retrospective PDF reweighting operations
using the LHAPDF\,6 API, as described in Section~\ref{sec:reweighting}, should
see particularly noticeable performance increases with\linebreak[100] LHAPDF\,6,
since the initial-state parton IDs are already known and hence only two parton
flavours need to be evolved per event.

\begin{table}[t]
  \centering
  \caption{Times taken for phase space integration and CKKW-merged event
    generation using the Sherpa MC event generator with LHAPDF\,5 ($t_5$) and LHAPDF\,6 ($t_6$)
    via interface code optimised for each LHAPDF version, and the speed improvement ratio
    $t_5/t_6$. In all cases LHAPDF\,6 runs faster than v5, in some (process- and PDF-specific)
    cases, faster by factors of 2--6.}
  \label{tab:sherpacpu}
  \begin{tabular}{lrrrr}
    \toprule
    Process/PDF & $t_5$ & $t_6$ & $t_5/t_6$\\
    \midrule
    \multicolumn{5}{l}{\textbf{Cross-section integrations, 1M phase space points}}\\
    \addlinespace
    \multicolumn{5}{l}{CT10}\\
    $pp \to jj$        & 23'10"  & 9'17" & 2.5\\
    $pp \to \ell\ell$  &  4'12"  & 2'02" & 2.1\\
    $pp \to H$ (ggF)   &  0'20"  & 0'15" & 1.3\\
    \addlinespace
    \multicolumn{5}{l}{NNPDF23nlo}\\
    $pp \to jj$        & 54'40"  & 9'28" & 5.8\\
    $pp \to \ell\ell$  &  8'06"  & 2'33" & 3.2\\
    $pp \to H$ (ggF)   &  0'25"  & 0'11" & 2.3\\
    \addlinespace
    \midrule
    \addlinespace
    \multicolumn{5}{l}{\textbf{CKKW event generation, 100k $pp \to \,\le 4\;\text{jet}$ events}}\\
    \addlinespace
    \multicolumn{5}{l}{CT10}\\
    Weighted     &   43'02" &   35'47" & 1.2\\
    Unweighted   & 5h04'39" & 4h30'26" & 1.1\\
    \multicolumn{5}{l}{NNPDF23nlo}\\
    Weighted     & 47'47"   &   27'20" & 1.7\\
    Unweighted   & 6h44'47" & 4h48'26" & 1.4\\
    \bottomrule
  \end{tabular}
\end{table}

For code which has not been rewritten to use the LHAPDF\,6 API, a performance
improvement may be implemented in a future LHAPDF\,6 version, explicitly adding
caching of positional interpolation weights between evolution calls, so that
consecutive evaluations at the same phase space point do not need to fully
recompute the PDF interpolation. In an extreme case all required PDF derivatives
at grid knots could also be pre-computed, similarly to how the knot point $\log
x$ and $\log Q^2$ are currently computed during \kbd{PDF} initialisation;
however, this would be likely to introduce a memory bottleneck in the
computation, and methods such as use of space-filling curves to optimise CPU
cache usage would add significant complication.

Additional CPU performance improvements are also being considered, in particular
use of vectorised (and currently CPU-architecture-specific) SSE or AVX
instructions for parallel interpolation of all flavours, or multiple
simultaneous PDF queries. Vectorisation works best when there are no conditional
branchings, hence re-engineering the spline interpolation to make best use of
vectorisation would involve removing the current \kbd{if}-branching used to
identify the edges of $Q$ subgrids and instead using extrapolated ``halos''
surrounding each subgrid. However, such an approach may have numerical
consequences, particularly in how the edges of the grid and hand-over to
extrapolation is handled, and will not be taken lightly. We welcome feedback on
the extent to which particle physics computations are CPU-bound by LHAPDF
interpolation.

Equivalent concerns apply to the possibility to use general-purpose graphical
processing units (GP\,GPUs) for vectorised PDF evolution; parallel evolution of
$\mathcal{O}(13)$ parton flavours would not justify the trade-off of extra GPU
code-complexity and platform-specificity.  An alternative use would be to
compute many points in parallel, but this is often not a natural use since many
applications are Markov Chains where the next step is conditional on the result
at the current one. It could benefit PDF reweighting, however, and should GPU
implementations of matrix element event generation codes become
prevalent\,\cite{Hagiwara:2009aq,Giele:2010ks} then it will be natural for
LHAPDF to support GPU operation. For the time being we prefer not to prematurely
optimise for use-cases which may not manifest.

Parallel execution at the multi-thread level, or across multiple processes with
shared read-only memory, may also be useful in PDF reweighting and does not have
the technical overhead of GPGPU programming. LHAPDF\,6 does not include any
specific mechanisms to interface with multi-core frameworks such as OpenMP or
MPI, but is largely safe to use with applications written to use them. Since
there is some global state for the global configuration and the \kbd{PDFSet}
objects created and returned by \kbd{LHAPDF::getPDFSet()}, LHAPDF\,6 is not
100\% thread-safe; but if all \emph{changes} to global and set-level
configuration are made before the concurrent block, then use of PDF querying
operations on \kbd{PDF} objects allocated locally to each thread should be safe.

A final, usually very minor, speed improvement has been seen in the
initialisation time of LHAPDF\,6 PDF members. Since members are now located in
individual files rather than within one large file for the whole set, random
access to a particular PDF no longer requires ``scrolling'' through the rest of
the file and loading the rest of the set's members. This speed improvement is
not usually noticeable because the time taken to load a PDF set in either LHAPDF
version is far less than one second. Some unusual applications may need to
reload PDFs from file very frequently, however, and for such situations we have
made use of a custom fast parser of numeric data from ASCII files, where the
speed-up is achieved by ignoring the possibility of wide-character types
(e.g. Unicode) which are implicitly handled by C++'s I/O stream types. This
optimisation makes LHAPDF\,6 loading of whole PDF sets as fast as in
LHAPDF\,5. A further speed-up at initialisation time, if really desired, can be
achieved by zipping the PDF set directories into \kbd{.zip} files -- this trick
is described in the next section, since the main effect is upon disk space
rather than significant speed improvements.

\subsection{Disk space requirements}

The disk space requirements of LHAPDF\,6 data sets are largely similar to those
of their LHAPDF\,5 equivalents. For example, the CT10nlo PDF set file is 21\;MB
in LHAPDF\,5 and the equivalent LHAPDF\,6 directory contains 33\;MB of data
files; showing the opposite trend, the NNPDF\,2.3 NLO PDFs are all typically
95\;MB in LHAPDF\,5 and 84\;MB in LHAPDF\,6.

LHAPDF\,6's use of directories and member-specific data files within does permit
an extreme disk-space optimisation where \kbd{.dat} files which will not be used
can be removed from the set directory. This is not recommended in typical usage,
but may be found to be helpful when e.g. the central members of many PDF sets
need to be available, but error sets are not needed at all. A less extreme
optimisation is to compress each \kbd{.dat} grid data file into \kbd{.dat.zip}
or \kbd{.dat.gz} file and use the \kbd{zlibc} library to access them as if they
were unzipped.  This can be done without modifying any code by (on Linux
systems) setting the \env{LD\_PRELOAD} environment variable to the path to
\kbd{zlibc}'s \kbd{uncompress.so} library, %  -- LHAPDF\,6 access to files within the
% subdirectories then proceeds seamlessly,
and the typical compression factor of 3--4 reduces the disk space needed to
store the data and can also speed up PDF initialisation. There are typically too
many portability issues with this approach to currently make zipped data files
standard in LHAPDF\,6, but the option exists for applications which need it.

% Trade-off with readability -- it can be useful to have easy access to .info
% files.

%% file: migration.tex
%auto-ignore

A major task, as substantial as writing the new library, has been the migration
of PDFs from the multitude of LHAPDF\,5 formats to the new \kbd{GridPDF} format
and interpolator, and then validating their faithfulness to the originals. This
has been done in several steps, starting with a Python script which used the
LHAPDF\,5 interface (with some extensions) to extract the grid knots and dump
the PDF data at the original knot points into the new format. This script has
undergone extensive iteration, as support was added for subgrids,
member-specific metadata, etc., and to allow more automation of the
conversion process for hundreds of PDFs.

The choice was made to only convert the most recent PDF sets in each family
unless there were specific requests for earlier ones: this collection is more
than 200 PDF sets, and only a few older PDFs have been requested in addition to
the latest sets.

\begin{figure*}[t]
  \centering
  \includegraphics[width=0.48\textwidth]{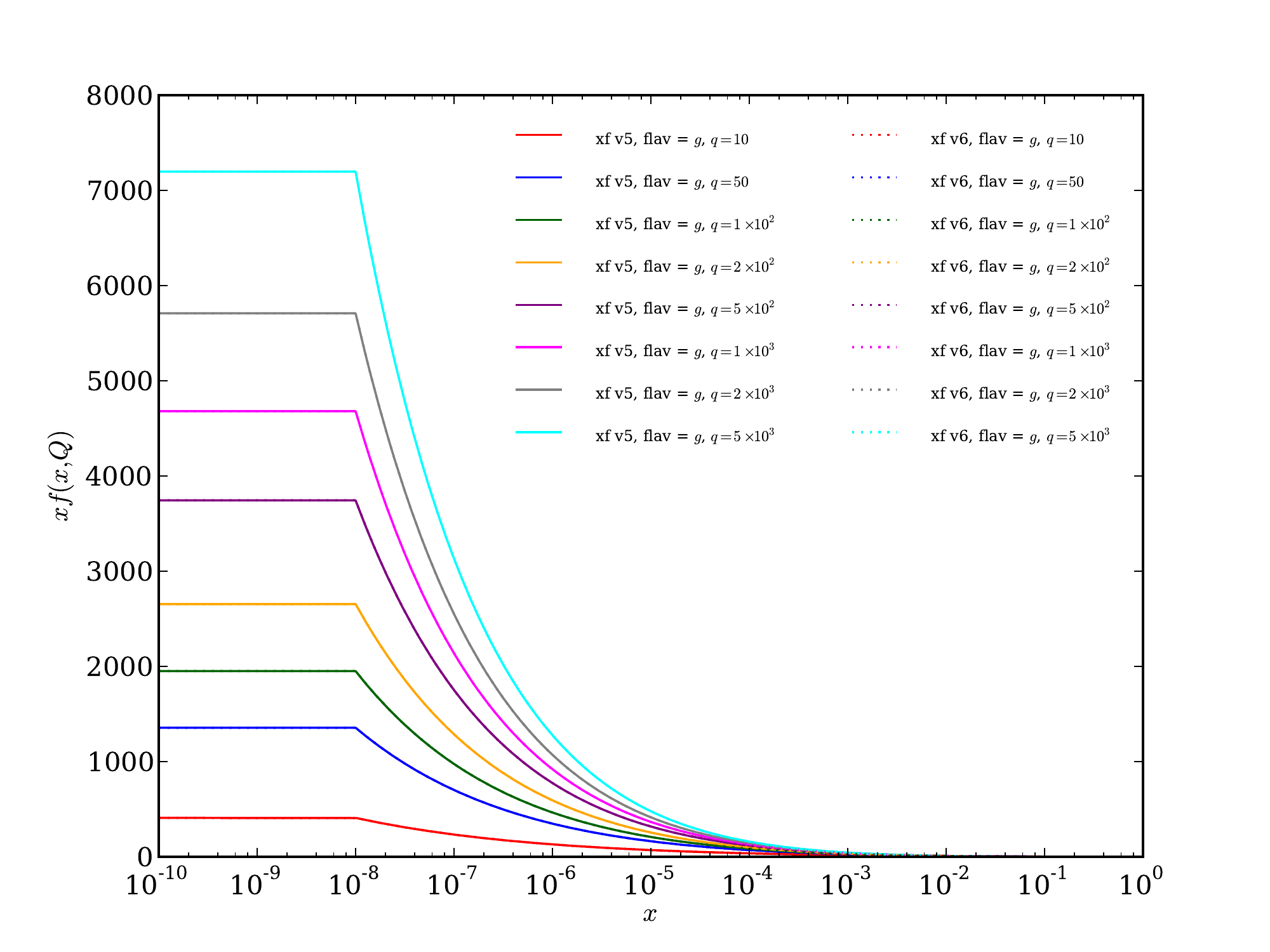}
  \includegraphics[width=0.48\textwidth]{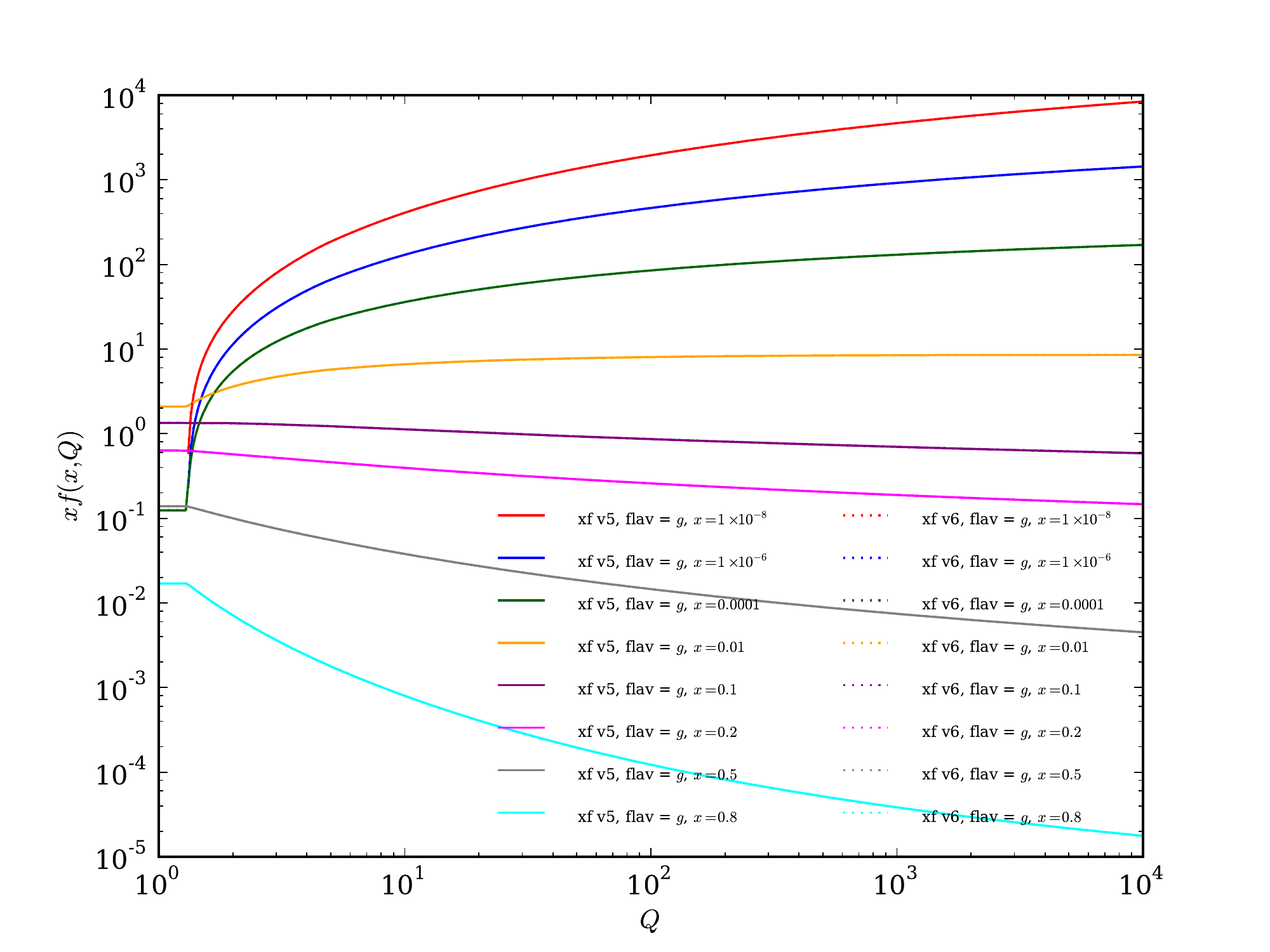}\\
  \includegraphics[width=0.48\textwidth]{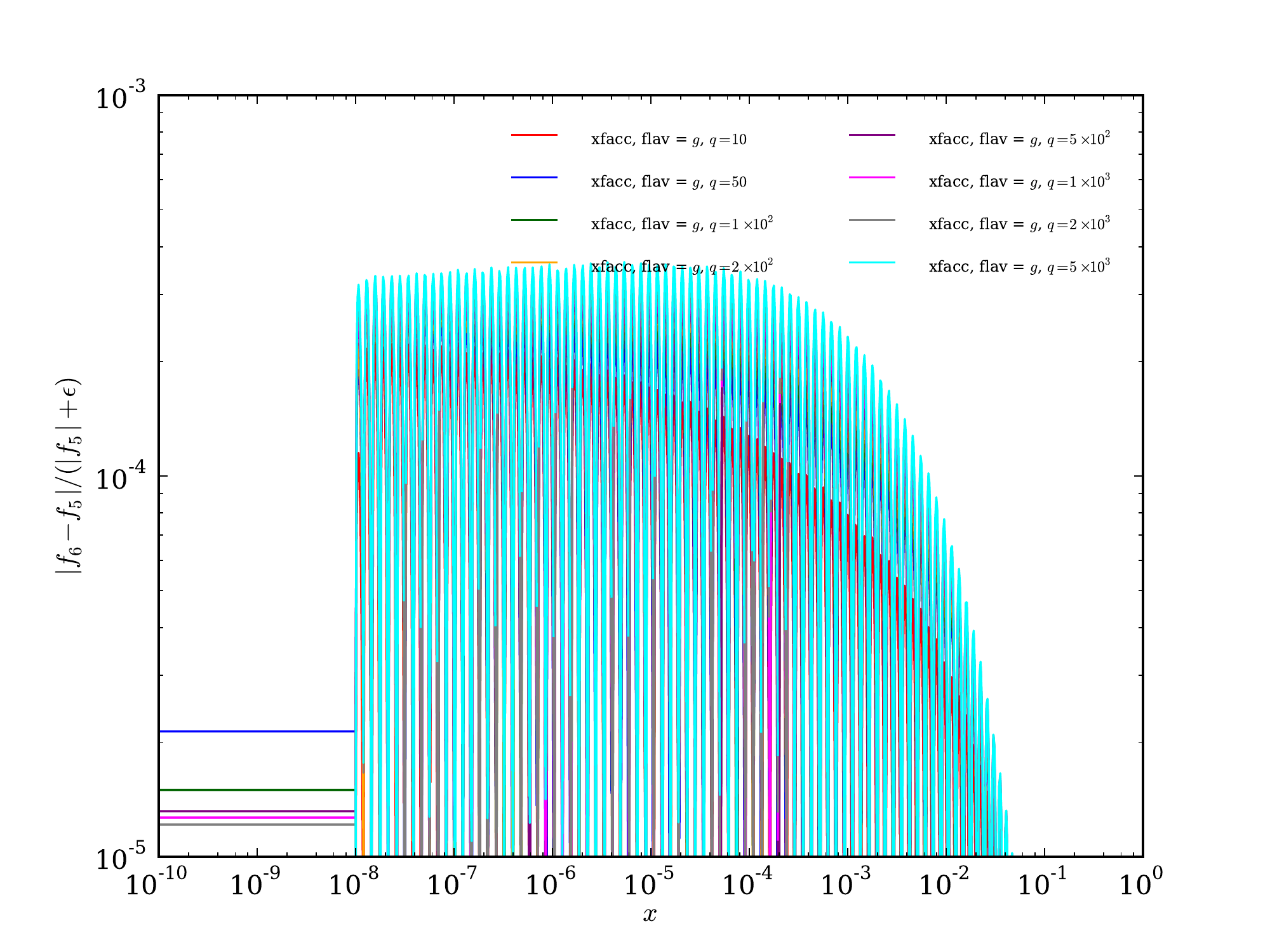}
  \includegraphics[width=0.48\textwidth]{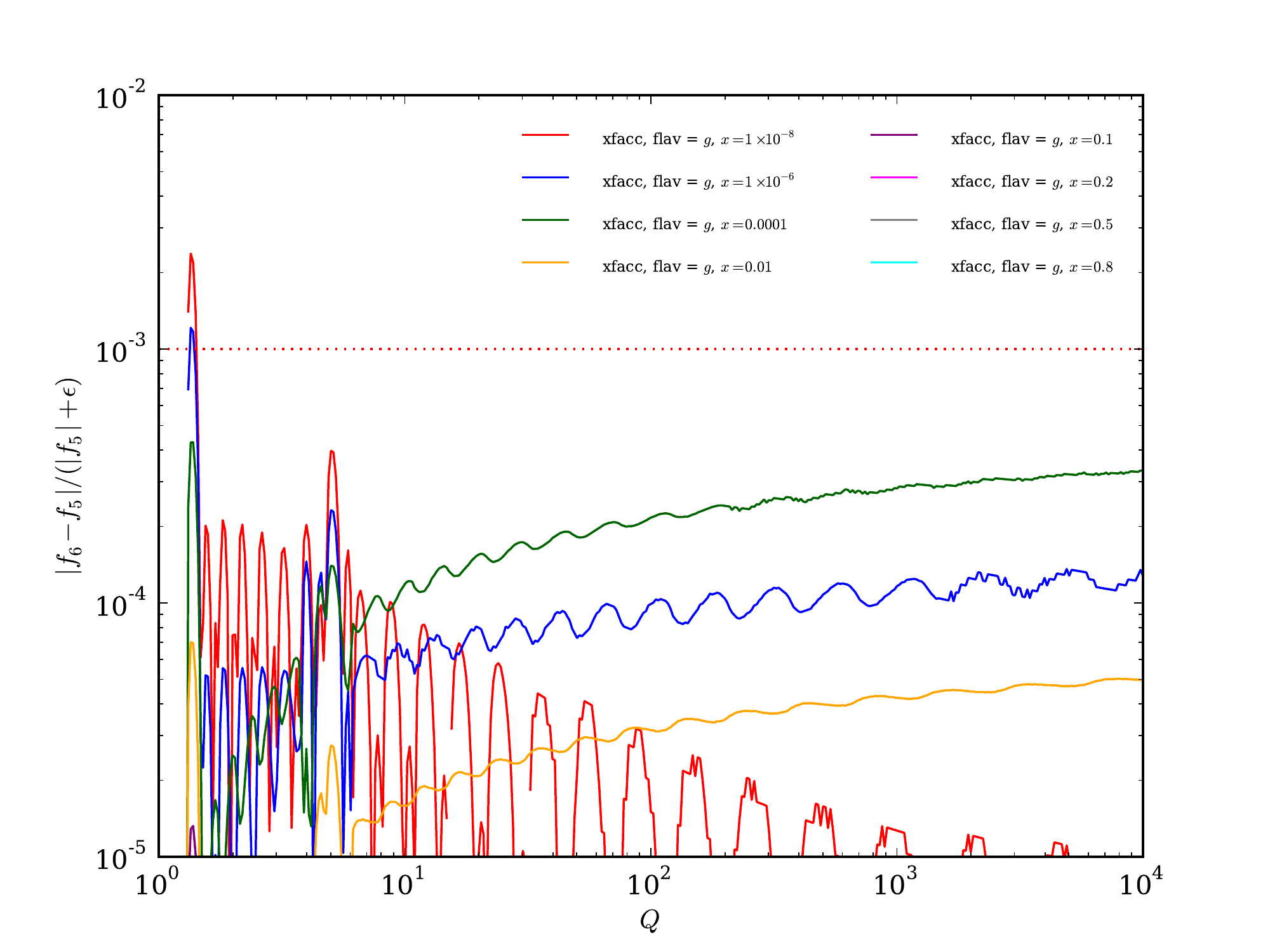}
  \caption{Example comparison plots for the validation of the
    CT10nlo\,\cite{Lai:2010vv} central gluon PDF, showing the PDF behaviour as a
    function of $x$ on the left and $Q$ on the right. The upper plots show the
    actual PDF shapes with both the v5 and v6 versions overlaid, and the lower
    plots contain plots of the corresponding v5 vs. v6 regularised accuracy
    metrics. The differences between v5 and v6 cannot be seen in the upper
    plots, since the fractional differences are everywhere below one part in
    1000 except right at the very lowest $Q$ point where the two PDFs freeze in
    very slightly different ways. The oscillatory difference structures arise
    from small differences in the interpolation between the identical
    interpolation knots.}
  \label{fig:ct10cmp}
\end{figure*}

To validate PDFs, a comparison system was developed, using a C++ code to dump
PDF $xf$ values in scans across $\log x$ and $\log Q$ (as well as \alphaS values
in $\log Q$) in the ranges $x \in [10^{-10}, 1]$ and $Q \in [1, 10^4]\;\GeV$,
with 10 sample points per decade in each variable. For scans in $x$, fixed
values of $Q \in \{10,\linebreak[1] 50,\linebreak[1] 100,\linebreak[1]
200,\linebreak[1] 500,\linebreak[1] 1000,\linebreak[1] 2000,\linebreak[1]
5000\}$\;\GeV were used, and for the scans in $Q$, fixed $x \in
\{10^{-8},\linebreak[1] 10^{-6},\linebreak[1] 10^{-4},\linebreak[1]
10^{-2},\linebreak[1] 0.1,\linebreak[1] 0.2,\linebreak[1] 0.5,\linebreak[1]
0.8\}$ were used. The same C++ code was used -- with some compile-time
specialisation -- to dump values from both LHAPDF\,5 and 6, to ensure exactly
equivalent treatment of the two versions.

The corresponding data files from each version were then compared to each other
using a difference metric which corresponds to the fractional deviation of the
v6 value from the original v5 one in regions where the $xf$ value is large, but
which suppresses differences as the PDFs go to zero, to minimise false
alarms. An ad hoc difference tolerance of $10^{-3}$ was chosen on consultation
with PDF authors as a level to which no-one would object, despite differences in
opinion on e.g. preferred interpolation schemes. This level, as illustrated in
Fig.~\ref{fig:ct10cmp} for the CT10nlo central PDF member validation, has been
achieved almost everywhere for the majority of PDFs. Several differences were
found this way, which helped with debugging the LHAPDF\,6 code, the migration
system, and occasionally the numerical stability of the original PDF's
interpolation grid.

Before being officially made available for download from the LHAPDF website and
AFS \& CVMFS locations, the validation plots resulting from this process had to
be checked by the original set authors as well as the LHAPDF\,6 team. To date
more than 200 PDF sets, from the ATLAS%\,\cite{atlaspdf}
, CTEQ \& CJ\,\cite{Lai:2010vv,ACCARDI:2013gra}, HERAPDF\,\cite{Radescu:2011cn},
MRST\,\cite{Martin:2004dh,Sherstnev:2007nd,Sherstnev:2008dm},
MSTW\,\cite{Martin:2009iq,Martin:2009bu,Martin:2010db}, and
NNPDF\,\cite{Ball:2011uy,Ball:2012cx} fitting collaborations, have been approved
in this way. In addition, over 100 new sets have been supplied directly to
LHAPDF in the new native data format from the
JR\,\cite{Jimenez-Delgado:2014ysa}, METAPDF\,\cite{Gao:2013bia},
MMHT\,\cite{Harland-Lang:2014zoa}, and %\linebreak[20]
NNPDF\,\cite{Ball:2014uwa} collaborations. Tools to help with PDF migration from
LHAPDF\,5 and validation of migrated or independently constructed PDFs may be
found in the \kbd{migration} subdirectory of the LHAPDF source package, but only
in the developers' version available from the Mercurial repository.

% PDFs which have been approved by their original authors are
% available in the ``main'' download area, while those yet to be blessed are
% contained in a subsidiary ``unvalidated'' staging area. Validated PDFs are also
% made available on CERN AFS, at
% \path{/afs/cern.ch/sw/lcg/external/lhapdfsets/current}.

%% file: summary.tex
%auto-ignore

After a lengthy public testing period, the first official LHAPDF\,6 version,
6.0.0, was released in August 2013. As described, this new version of LHAPDF
maintains compatibility with applications written to use the\linebreak[20] LHAPDF\,5 code
interfaces, while providing much more powerful models for dynamic allocation of
PDF memory and for parton density metadata.

The new design also provides a unified data format and routines for PDF
interpolation, which decouples new releases of PDF sets from the slower release
cycle of the LHAPDF software library. The new design which allows very general
parton content has also proven useful for the new generation of NNPDF sets which
include polarised partons and photon
constituents~\cite{Nocera:2014gqa,Ball:2013hta}, and for implementing
fragmentation functions using the PDF interpolation machinery. %~\cite{nnpdf,Bauer:2013bza}.
Several PDF sets have already been supplied directly to the LHAPDF\,6 library in
the new native format, which simplifies and speeds up the release of new PDFs for
PDF users and authors alike.

The new code design vastly reduces the memory requirements of the library
compared to the several GB demanded by LHAPDF\,5, meaning that it can
efficiently use multiple full PDF sets at the same time -- a task which was
unfeasible with Grid-distributed builds of LHAPDF\,5. Gains in CPU performance,
although a smaller effect than the fix to LHAPDF\,5's pathological memory
requirements, are also possible with the new structure due to the ability to
interpolate single flavours at a time rather than being forced to always evolve
all of a PDF's constituent flavours at the same time: this particularly improves
performance in reweighting applications where at most two parton flavours need
to be evolved per event. There is room for further CPU performance improvements
by adding explicit caching of some interpolation coefficients at a given $(x,
Q^2)$ point, and with more work the code can be optimised to allow use of
vectorised CPU instructions. Addition of flavour aliasing or compressed data
file reading could reduce the data size on disk. However, all such performance
optimisations need to be judged according to the real-world benefits which they
offer, against the code complexity which they typically introduce.

Finally, LHAPDF\,6 provides new tools for PDF uncertainty and reweighting
calculations, to respond to the increasingly complex ways in which particle
physics experiment and phenomenology use PDFs.

% \subsection{Limitations and prospects}

At present the scope of LHAPDF\,6 is intentionally more LHC-focused than
LHAPDF\,5. Accordingly, no QCD evolution is planned for the library since this
functionality ended up virtually unused in LHAPDF\,5. Several quality external
libraries\,\cite{Botje:2010ay,Salam:2008sz,Bertone:2013vaa} exist to perform this evolution
and generate the grid files -- or if desired, the \kbd{PDF} class can be derived
from to call an evolution library at runtime. Similarly, there is at present no
plan to support resolved virtual photon structure functions or
transverse-momentum-dependent PDFs, which require additional parameters in the
interpolation space. % nor double-parton PDFs

Nuclear corrections to nucleon PDFs are also not currently supported in a
transparent way, but this is planned for a near-future LHAPDF version. In the
meantime, external nuclear correction factors such as the EPS
sets\,\cite{Eskola:2009uj} can be applied explicitly to nucleon PDFs from
LHAPDF. Nuclear PDFs with the corrections already ``hard-coded'' into LHAPDF\,6
grids are also trivially supported, since these are indistinguishable from
nucleon PDFs, other than via the \kbd{Particle} metadata key which can declare
the nucleus/ion as the parent particle in place of the usual proton -- this is
another strength of the decision to use the standard PDG particle ID number
scheme in LHAPDF\,6.

In summary, LHAPDF\,6 is fully operational at the planned level, offers very
significant improvements in performance and capabilities over LHAPDF\,5, and is
recommended as the production version of LHAPDF in the high-precision era of
collider physics which begins with LHC Run\,2.